\documentclass[prb,twocolumn,aps,amsmath,amssymb]{revtex4}
\usepackage[T1]{fontenc}
\usepackage[utf8]{inputenc}
\usepackage{graphicx}
\usepackage{amsmath}
\usepackage{amssymb}
\usepackage{dcolumn}
\usepackage{float}
\usepackage{bm}
\usepackage{amsfonts,times}
\usepackage{xcolor}
\usepackage[breaklinks=true,colorlinks,citecolor=blue,linkcolor=blue,urlcolor=blue]{hyperref}

\DeclareMathAlphabet{\bi}{OML}{cmm}{b}{it}

\def\be{\begin{equation}}
\def\ee{\end{equation}}
\def\bearr{\begin{eqnarray}}
\def\eearr{\end{eqnarray}}

\begin{document}
\title{Tunable persistent currents in a spin-orbit coupled pseudospin-1 fermionic quantum ring}
\author{Mijanur Islam}
\author{Saurabh Basu}

\affiliation{Department of Physics, Indian Institute of Technology-Guwahati, Guwahati-781039, India.}
\normalsize

\begin{abstract}
We conduct a thorough study of the persistent currents in a spin-orbit coupled $\alpha$-$T_3$ pseudospin-1 fermionic quantum ring (QR) that smoothly interpolates between graphene ($\alpha=0$, pseudospin-1/2) and a dice lattice ($\alpha=1$, pseudospin-1). In particular, we have considered both intrinsic spin-orbit coupling (ISOC) and Rashba spin-orbit coupling (RSOC) in addition to an external magnetic field, and have systematically enumerated their individual and combined effects on the charge, valley and the spin-polarized persistent currents. The energy levels of the system comprise of the conduction bands, valence bands and flat bands which show non-monotonic dependencies on the ring radius, $R$ of the QR, in the sense that, for small $R$, the energy levels vary as $1/R$, while the variation is linear in $R$ for large $R$. The cases corresponding to zero magnetic field are benchmarked with those for finite external fields. Further, it is noted that the flat bands demonstrate dispersive behaviour, and hence can contribute to the transport properties only when ISOC is non-zero. Moreover, the RSOC yields spin-split bands, thereby contributing to the spin-resolved currents, together with distinct degeneracies for different spin branches. The persistent currents in the charge, valley, and spin sectors for each of these cases oscillate as a function of the magnetic field with a period equal to the flux quantum, as they should be, and depend upon the spin-orbit coupling terms. Further, we have explored the role played by the parameter $\alpha$ in our entire analysis to ascertain the effect of the flat bands.  
\end{abstract}

\maketitle

\section{Introduction}
Recent research on electronic properties in quantum confined nanoscale systems, such as quantum wells, quantum wires, quantum dots, and quantum rings (QRs) has seen significant progress, revealing new phenomena and their potential for device applications. Studies on metallic and semiconducting rings have received considerable theoretical and experimental attention over the years [\onlinecite{But,Che,Che1,Che2,Le,Mon,Cha,Avi,Bou,Mai,Lor,Alf,Fu,Cli,Per,Qi,Chav,Na}]. Moreover, the role of magnetic field on the transport properties of such systems have remained at the center of attention. The energy levels of the ring and its properties as a function of the magnetic field are studied thoroughly [\onlinecite{Vie}]. The investigation of quantum rings at a mesoscopic scale in the presence of external magnetic fields has unveiled intriguing physical phenomena, such as the Aharonov-Bohm effect [\onlinecite{Aha,Key,Pee}], Aharonov-Casher effect [\onlinecite{Aha1,Ber,Joi}], magnetic oscillations [\onlinecite{Mei}], persistent currents [\onlinecite{Avi,Bou,Schi,Tan}],  many-body correlations [\onlinecite{Key1}], and spin-orbit induced Berry phases [\onlinecite{Mor}]. In the presence of a magnetic field, when a ring is threaded a magnetic flux $\Phi$, it becomes possible to eliminate the vector potential from the Schrödinger equation by implementing a gauge transformation. This leads to a modification of the boundary condition as follows, $\psi_n(x + L) = e^{(2\pi i\Phi/\Phi_0)} \psi_n(x)$, where $L$ represents the circumference of the ring. This situation can be likened to the one-dimensional Bloch problem, wherein the Bloch wave vector $k_n$ is defined as $(2\pi/L)\Phi/\Phi_0$. Consequently, the energy levels $E_n$ and other related physical properties exhibit periodic behavior with respect to $\Phi_0$ [\onlinecite{But,Che,Che1}]. In simple language, it means that the Hamiltonian remains invariant as flux is varied in integer multiples of the flux quantum. Quantum confinement within the ring structure not only leads to the coupling of charge carriers to an effective gauge field but also significantly influences coherent-electron interference [\onlinecite{Lan,Ric,Ple}]. Quantum rings also hold promise for diverse applications, such as terahertz detectors [\onlinecite{Hua1}], efficient solar cells [\onlinecite{Wu}] and memory devices through electrically tunable optical valves [\onlinecite{You}] and single-photon emitters [\onlinecite{War,Abb}]. Additionally, with consideration of spin-orbit interactions, particularly of Rashba type [\onlinecite{FE}], semiconductor QRs exhibit a variety of fascinating spin phenomena [\onlinecite{Ric,Spin_dep1,Spin_dep2,Spin_dep4,Spin_dep5,Spin_dep6}].

Arguably, the most promising material of the century, graphene [\onlinecite{Graph_exp1,Graph_exp2,Graph_exp3,Graph_exp4}] has garnered significant attention in the study of quantum ring systems, both theoretically and experimentally. This is mainly due to its unique properties, such as the involvement of linearly dispersive 'massless' Dirac fermions  [\onlinecite{Zh,De}], possible topological phases originating from the violation of time reversal symmetry [\onlinecite{Haldane}], the manifestation of Aharonov-Bohm oscillations in the presence of a magnetic field  [\onlinecite{Grap_Lith1,Grap_Lith2,Grap_Lith3,PR,Fa}] etc. In the graphene QR, the persistent currents are induced by the breaking of time-reversal symmetry [\onlinecite{PR}]. Additionally, the confinement of electrons in the ring structure leads to the controlled lifting of valley degeneracy in the presence of a magnetic field [\onlinecite{Bol,DS}]. Studies have also shown evidence of broken valley degeneracy [\onlinecite{PR,DR}] and an interplay between valley polarization and electron-electron interaction [\onlinecite{DS}] in graphene quantum rings. Numerous investigations have been conducted in recent years to understand the microscopic details of graphene QRs under external magnetic fields, with and without invoking spin-orbit interactions  [\onlinecite{Graph_Numr1,Graph_Numr3,Graph_Numr4,Graph_Numr5,Graph_Numr6,Graph_Numr7,
Graph_Numr9,Graph_Numr10,Graph_Numr11,Graph_Model1,Graph_Model2,
Graph_Bilayer,hybrid_grapR}].  It is demonstrated that a graphene QR has potential application in future optoelectronic [\onlinecite{Graph_Opto}] and interferometric [\onlinecite{Graph_Interf}] devices.

 Furthermore, there exists an interesting variant of the honeycomb structure of graphene with $T_3$ symmetry, namely $\alpha$-$T_3$ lattice [\onlinecite{MI,Su,Vi}]. The low-energy massless excitations of the $\alpha$-$T_3$ lattice near the Dirac points consists of three branches, two of them being linearly dispersive, while the other is a non-dispersive flat band. The $\alpha$-$T_3$ lattice is defined as a special honeycomb like structure with an additional inequivalent site present at the center of each hexagonal honeycomb lattice of graphene structure. That additional site is connected only with one of the two inequivalent sites of the honeycomb lattice with a distinct (with respect to the others) hopping amplitude. The ratio of the hopping amplitudes between these two hopping amplitudes is $\alpha$. With the continuous tuning of the parameter $\alpha\in(0,1)$, the $\alpha$-$T_3$ lattice provides an interpolation between the honeycomb structure of graphene ($\alpha=0$) and the dice lattice ($\alpha=1$). A dice lattice can be realized by growing trilayers of cubic lattices (e.g., SrTiO$_3$/SrIrO$_3$/SrTiO$3$) in the (111) direction [\onlinecite{Wa}]. Further, in the context of cold atom, a suitable arrangement of three counterpropagating pairs of laser beams  can produce an optical dice lattice [\onlinecite{Ur}]. Moreover, the electronic dispersion of Hg$_{1-x}$Cd$_x$Te quantum well at a certain critical doping can also be mapped onto the $\alpha$-$T_3$ model in the intermediate regime (between dice and graphene), corresponding to a value of $\alpha=1/\sqrt{3}$ [\onlinecite{Mal}], where the band structure comprises of linearly dispersing conduction and valence bands, in addition to a flat band. The pseudospins of graphene and $\alpha$-$T_3$ lattices are fundamentally different, while the former has pseudospin $S=\frac{1}{2}$, whereas the quasiparticles of the latter system obey the Dirac-Weyl equation with pseudospin $S = 1$. A plethora of studies have been performed in recent years to probe various equilibrium[\onlinecite{T3_Hall1, T3_Hall2, Klein2, Weiss, ZB, Plasmon1, Plasmon2, Plasmon3, Plasmon4, Mag_Opt1, Mag_Opt2, Mag_Opt3, Mag_Opt4, RKKY1, RKKY2, Min_Con, Ghosh_Topo, spin_hall, aT3_Mijanur}] as well as nonequilibrium[\onlinecite{Bashab1, Bashab2, Iurov_Floq, Mojarro, TB_Floq, Trans_aT3, TB_Trans}] properties of the $\alpha$-$T_3$ lattice. On the other hand, topological phases in a Haldane dice [\onlinecite{Ghosh_Topo,Sayan_dice}] lattice model as well as Rashba dice model [\onlinecite{Wa,Rahul1,Rahul2}] have invoked intense attention in the recent years. Furthermore, the quantum spin Hall phase transition [\onlinecite{QSH_alpha}] in the $\alpha$-$T_3$ lattice has also observed.
 
 Motivated by the promising prospects of the quantum rings, in this paper we present analytical results for the energy levels of an ideal ring of pseudospin-1 $\alpha$-$T_3$ lattice. We consider the bare Dirac Hamiltonian, augmented with either the Rashba or the intrinsic spin-orbit coupling (SOC) terms, as well as focus on the Kane-Mele model on a ring that includes the Rashba spin-orbit term and complex next nearest neighbour (NNN) hopping and can be considered as two copies of the Haldane model with different Haldane fluxes for the two spins. The Haldane term is often referred to as the intrinsic SOC. The Rashba SOC term which obeys all the symmetries that a parent $\alpha$-$T_3$ ring has can be controlled by an external electric field [\onlinecite{Kane}] that breaks the mirror symmetry with respect to the $\alpha$-$T_3$ plane, while the intrinsic coupling can be enhanced through edge heavy-atom functionalization [\onlinecite{Func}]. The above discussion necessitates a thorough study of an $\alpha$-$T_3$ QR with the intrinsic and the Rashba SOCs and further, to reflect on the tunability of the persistent current, we shall additionally include an external magnetic field as well.

\begin{figure}[h!]
\centering
\includegraphics[width=8.5cm, height=9.7cm]{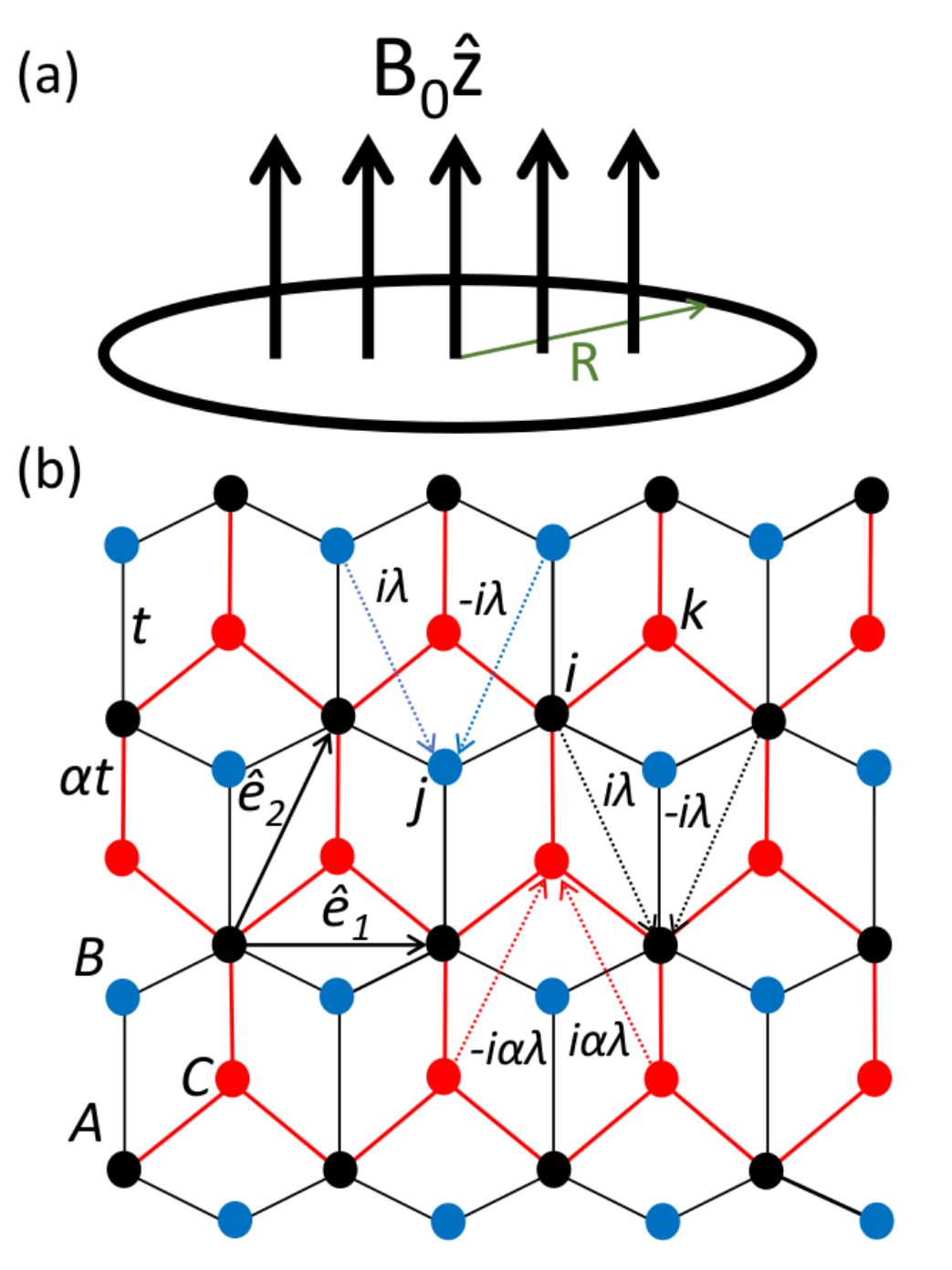}
\caption{(Color online) (a) Schematic diagram of the $\alpha$-$T_3$ ring of radius $R$ subjected to a perpendicular magnetic field $B=B_0\hat{z}$. (b) Lattice structure of the $\alpha$-$T_3$ lattice. Here, A, B, and C lattice sites are shown by black, blue, and red dots respectively. The dashed arrows represent the hopping among the next nearest neighbouring (NNN) atoms, such as B-A-B or C-A-C or A-B-A or A-C-A. $\hat{e_1}$ and $\hat{e_2}$ are the lattice unit vectors.}
\label{fig:ring_geo_soi}
\end{figure}
 
We start by obtaining a ring Hamiltonian of radius $R$ in polar coordinate [\onlinecite{FE,DR}]. We handle subtleties regarding the hermiticity of the Hamiltonian and exclude additional complexities arising from the edge effects for the finite QR. Consequently, we obtain the analytical expressions for the energy spectrum and the corresponding wave functions as a function of the ring radius and an external magnetic field. Particularly, we intend to see how the energy spectrum evolves as we approach the dice limit ($\alpha=1$) starting from the graphene ($\alpha=0$). However, our results with $\alpha=0$ i.e., graphene exactly match with the previously reported analytical and numerical results by the other groups [\onlinecite{Bol}]. Further, the inclusion of an external field would facilitate studies on the evolution of the spectral properties, the persistent current, and the interplay of the Aharonov-Bohm effect. Notably, we can tune the charge persistent currents, valley currents, and spin equilibrium currents via the parameters, $\alpha$, the strength of the Rashba and the intrinsic SOCs, allowing an interplay of these parameters and measurable quantity such as the currents.

A brief note on the choice of the spin-orbit coupling parameters will add a motivation to the results presented below. Through tight-binding calculations, it has been demonstrated that the numerical magnitude of ISOC is approximately 10 to 20 times smaller than that of RSOC [\onlinecite{Just1}]. Moreover, recent research has indicated that in cases where the both couplings are present and have comparable magnitudes, a mixing of the radial state occurs. Conversely, when one of these couplings dominates, the system exhibits non-overlap of the radial states [\onlinecite{Just2}]. Several investigations have been conducted on the pseudopin-1 $\alpha$-$T_3$ and dice lattice, utilizing an ISOC strengths within the range varying between $0.01t$ and $0.1t$ [\onlinecite{Ghosh_Topo,Sayan_dice,QSH_alpha}], and RSOC strengths within a range of roughly $0.1t$ to $t$ [\onlinecite{Wa, Rahul1, Rahul2}]. Additionally, the typical value for the nearest-neighbor hopping parameter in graphene is $t = 2.7eV$ [\onlinecite{Just3}]. In our calculations, we shall adopt $t = 1eV$ everywhere, and to explore the effects of ISOC and RSOC, we shall employ two distinct sets of parameter values, namely, ISOC to have values $0.05t, 0.1t$, and RSOC to be $0.5t, 0.8t$.
 
The organization of the paper is as follows. In Sec. \ref{Hal}, we discuss the various properties of pseudospin-1 $\alpha$-$T_3$ QR in presence of intrinsic SOC including the energy spectrum and the persistent currents. Similar results for the Rashba SOC are given in Sec. \ref{Rash}. In Sec. \ref{Kane}, we present the results for a Hamiltonian along with both the intrinsic and the Rashba SOCs, including the energy spectrum as well as the persistent currents of the ring. We summarize our results in Sec. \ref{Sum}. To give a structure to the preceding discussion we have included results in the absence and presence of an external magnetic field for each of the cases in different subsections. The energy level diagrams are presented as a function of the radius of the QR in presence and absence of the external field. Physical quantities such as the persistent currents are also investigated as a function of the external flux.

\section{Pseudospin-1 QR with ISOC}
\label{Hal}
We consider a ring of radius R, in the x-y plane, made from the pseudospin-1 $\alpha$-$T_3$ lattice as depicted in Fig. \ref{fig:ring_geo_soi}(a). The $\alpha$-$T_3$ lattice structure is schematically shown in Fig. \ref{fig:ring_geo_soi}(b). The hopping energy between the A and B sites is $t$, while the hopping energy between the sites A and C is $t^\prime=\alpha t$. Including the intrinsic spin-orbit  coupling, the $\alpha$-$T_3$ lattice Hamiltonian is given by [\onlinecite{Haldane,QSH_alpha}],

\begin{widetext}
\begin{equation}
\label{soc}
H = \sum_{\langle ij \rangle} tc_{i\sigma}^\dagger c_{j\sigma}+\sum_{\langle ik\rangle}t^\prime c_{i\sigma}^\dagger c_{k\sigma}+\frac{i\lambda}{3\sqrt{3}}\sum_{\langle \langle ij\rangle\rangle\sigma\sigma^\prime}\nu_{ij}c^\dagger_{i\sigma}\sigma_zc_{j\sigma^\prime}+\frac{i\lambda^\prime}{3\sqrt{3}}\sum_{\langle \langle ik\rangle\rangle\sigma\sigma^\prime}\nu_{ik}c^\dagger_{i\sigma}\sigma_zc_{k\sigma^\prime}
\end{equation}
\end{widetext}
where $c^\dagger_{i,j,k}(c_{i,j,k})$ is the creation (annihilation) operator of electrons on the corresponding A, B, and C sites denoted by $i, j, k$ indices, respectively. The first term is the electron hopping between the A and B sites, while the second one is that between the A and C sites. The summation of $\langle ij\rangle$ $(\langle ik \rangle)$ runs over the nearest neighbour (NN) sites A-B (A-C). The third term is the next nearest neighbour (NNN) hopping of electrons $\langle \langle ij\rangle\rangle$ (double angular brackets denote the NNN hopping are to distinguished from the single ones that stand for NN hopping) between A-B-A or B-A-B representing the spin-orbit coupling proposed by Kane and Mele in graphene [\onlinecite{Kane}], $\lambda$ is the SOC strength. $\nu_{ij}$ $(\nu_{ik})=1$ for the NNN hopping to be anticlockwise and $\nu_{ij}$ $(\nu_{ik})=-1$ (depending on the orientation of the two nearest neighbour bonds $\bf{d_1}$ and $\bf{d_2}$ the electron traverses in going from site j to i) if it is clockwise with respect to the positive $z$ axis (which is perpendicular to the lattice $xy$ plane), $\sigma_z, \sigma$, and $\sigma^\prime$ are the real spin Pauli operators. The last term describes the extra SOI due to the introduction of the hub C atoms into the graphene lattice, i.e., the C-A-C and A-C-A NNN hoppings are feasible (B-C-B and C-B-C are neglected). $\lambda^\prime$ is the corresponding SOI strength and set as $\lambda^\prime = \alpha \lambda$ (similar to $t^\prime = \alpha t$), so that $\alpha=0$ reproduce all the results for graphene.

In the Bloch representation, the above lattice Hamiltonian can be transformed into a continuum one. Therefore, the total low-energy Hamiltonian around the Dirac points are governed by a pseudospin-1 Dirac-Weyl Hamiltonian of the form,
\begin{widetext}
\begin{equation}
\label{eqn:soi_Ham}
H(\sigma)=\hbar v_F\begin{pmatrix}
0 & (\zeta q_x-iq_y)\cos\xi & 0\\
(\zeta q_x+iq_y)\cos\xi & 0 & (\zeta q_x-iq_y)\sin\xi\\
0 & (\zeta q_x+iq_y)\sin\xi & 0
\end{pmatrix}\\
-\zeta\sigma\tilde{\lambda}\begin{pmatrix}
\cos\xi & 0 & 0\\
0 & \sin\xi-\cos\xi & 0\\
0 & 0 & -\sin\xi
\end{pmatrix}
\end{equation}
\end{widetext}
where $\tan\xi=\alpha$, $q_x$ and $q_y$ donate the in-plane mechanical momentum operator, $\zeta=\pm 1 = {\bf K}$ $({\bf K^\prime})$ representing the {\bf K} ({\bf K$^\prime$}) valley, $\tilde{\lambda}=\lambda/\cos\xi$, and $\hbar v_F=\sqrt{3}at/2\cos\xi$, while $\sigma$ denotes the real spin degrees of freedom. The Hamiltonian in Eq. (\ref{eqn:soi_Ham}) can be expressed in the polar ($r, \theta$) coordinates as,
\begin{widetext}
\begin{equation}
\label{43}
H_{ring}(\sigma)=\begin{pmatrix}
-\zeta\sigma\tilde{\lambda}\cos\xi & \hbar v_Fe^{-i\theta}(-i\zeta\frac{\partial}{\partial r}-\frac{1}{r}\frac{\partial}{\partial\theta})\cos\xi & 0\\
\hbar v_Fe^{i\theta}(-i\zeta\frac{\partial}{\partial r}+\frac{1}{r}\frac{\partial}{\partial\theta})\cos\xi & \zeta\sigma\tilde{\lambda}(\cos\xi-\sin\xi) & \hbar v_Fe^{-i\theta}(-i\zeta\frac{\partial}{\partial r}-\frac{1}{r}\frac{\partial}{\partial\theta})\sin\xi \\
0 & \hbar v_Fe^{i\theta}(-i\zeta\frac{\partial}{\partial r}+\frac{1}{r}\frac{\partial}{\partial\theta})\sin\xi & \zeta\sigma\tilde{\lambda}\sin\xi\\\end{pmatrix}
\end{equation}
\end{widetext}
The eigenstates of $H$ are obtained as,
\begin{equation}
\label{Eq:wave_soi}
\Psi_\sigma (r,\theta)=\begin{pmatrix}
\Xi_{1\sigma}(r)e^{i(m-\zeta)\theta}\\
\Xi_{2\sigma}(r)e^{im\theta}\\
\Xi_{3\sigma}(r)e^{i(m+\zeta)}\theta
\end{pmatrix},
\end{equation}
where the integer $m$ labels the orbital angular momentum quantum number and $\Xi_{i\sigma}$ denotes the amplitudes corresponding to the three sublattices. We follow the earlier approaches [\onlinecite{FE,Spin_dep1,Spin_dep2,Bol,DR,Graph_Model1, Graph_Model2,Mijanur}] employed for an ideal one dimensional quantum ring by freezing the radial part, $r=R$ in the eigensolutions. Thus, in this case of an ideal ring of radius $R$, the momentum of the carriers along the radial direction is zero. This turns out to be an obvious choice in constructing a hermitian Hamiltonian one should make the replacement $\frac{\partial}{\partial r}=-\frac{1}{2R}$. Therefore, the Hamiltonian corresponding to an ideal $\alpha$-$T_3$ ring is given by,
\begin{widetext}
\begin{equation}
\label{Eq.isoc_Ham}
H_{ring}(\sigma)=\begin{pmatrix}
-\zeta\sigma\tilde{\lambda} \cos\xi & -i\frac{\hbar v_F}{R} (m-\zeta/2) \cos\xi & 0\\
i\frac{\hbar v_F}{R} (m-\zeta/2)\cos\xi & \zeta\sigma\tilde{\lambda}\cos\xi - \zeta\sigma\tilde{\lambda}\sin\xi & -i\frac{\hbar v_F}{R} (m+\zeta/2)\sin\xi\\
0 & i\frac{\hbar v_F}{R} (m+\zeta/2) \sin\xi & \zeta\sigma \tilde{\lambda} \sin\xi\\
\end{pmatrix}
\end{equation}
\end{widetext}

\begin{widetext}

\begin{figure}[h!]
\centering
\includegraphics[width=15cm, height=9.5cm]{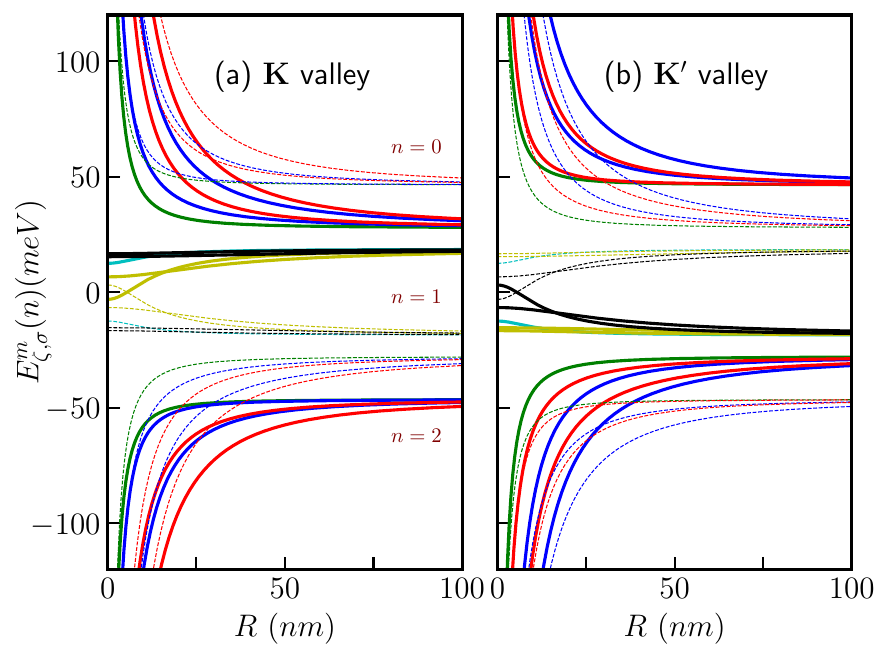}
\caption{(Color online) The spin-resolved band structures $E_{\zeta,\sigma}^m(n)$ vs $R$ of the $\alpha$-$T_3$ quantum ring for the (a) ${\bf K}$-valley and  (b) ${\bf K^\prime}$-valley in absence of an external field. $n=0,1,2$ denotes the CB, FB, and VB respectively. The solid and the dotted lines in each panel are for spin-$\uparrow$ and spin-$\downarrow$ bands respectively. The parameters are taken as $\tilde{\lambda}=0.05t$, $t=1eV$, and $\alpha=0.4$.}
\label{fig:soc_alphap4}
\end{figure}
\end{widetext}

The energy eigenvalues are obtained as,
\begin{equation}
\label{Energy_soc}
E_{\zeta,\sigma}^m(n) = 2\sqrt{-\frac{P}{3}}\cos\left[\frac{1}{3}\cos^{-1}\left(\frac{3Q}{2P}\sqrt{-\frac{3}{P}} \right)-\frac{2\pi n}{3} \right]
\end{equation}
where $n$ = 0, 1, and 2 are associated with the conduction band (CB), the flat band (FB), and the valence band (VB), respectively. Here $P$ and $Q$ are given by, 
\begin{align*}
P=\frac{\tilde{\lambda}^2}{2}\sin2\xi-\tilde{\lambda}^2-\frac{\hbar^2v_F^2}{R^2}\bigg[N^2\sin^2\xi+M^2\cos^2\xi\bigg]
\end{align*}
and
\begin{multline*}
Q =\frac{\zeta\sigma\tilde{\lambda}}{2}\sin2\xi\big[\tilde{\lambda}^2(\cos\xi-\sin\xi)+\frac{\hbar^2v_F^2}{R^2}\\
\big(M^2\cos\xi-N^2\sin\xi\big)\big]
\end{multline*}
respectively. The normalized spinor wavefunctions are given by,

\begin{widetext}

\begin{equation*}
\Psi^{m\zeta}_{n,\sigma}(R,\theta)=N_{n,\sigma}^\zeta\, e^{im\theta}\begin{pmatrix}
-i\frac{\hbar v_F}{R} M(E_{\zeta,\sigma}-\zeta\sigma\tilde{\lambda}\sin\xi) \cos\xi\, e^{-i\zeta\theta}\\
(E_{\zeta,\sigma}+\zeta\sigma\tilde{\lambda}\cos\xi)(E_{\zeta,\sigma}-\zeta\sigma\tilde{\lambda}\sin\xi)\\
i\frac{\hbar v_F}{R} N(E_{\zeta,\sigma}+\zeta\sigma\tilde{\lambda}\cos\xi) \sin\xi\, e^{i\zeta\theta}
\end{pmatrix},
\end{equation*}
with,
\begin{equation}
\label{WaveFn}
N_{n,\sigma}^\zeta=\frac{1}{\sqrt{\frac{\hbar^2v_F^2}{R^2}\Big[M^2(E_{\zeta,\sigma}-\zeta\sigma\tilde{\lambda}\sin\xi)^2 \cos^2\xi+N^2(E_{\zeta,\sigma}+\zeta\sigma\tilde{\lambda}\cos\xi)^2 \sin^2\xi\Big]+(E_{\zeta,\sigma}+\zeta\sigma\tilde{\lambda}\cos\xi)^2(E_{\zeta,\sigma}-\zeta\sigma\tilde{\lambda}\sin\xi)^2}}.
\end{equation}
\end{widetext}
where $M=(m-\zeta/2)$ and $N=(m+\zeta/2)$. One can easily verify that, when there is no SOC, that is, $\tilde{\lambda}=0$, the eigenvalues of Eq. (\ref{Energy_soc}) can be directly simplified onto the eigenvalues of the parent $\alpha$-$T_3$ QR with one zero energy mode [\onlinecite{Mijanur}]. When $\tilde{\lambda}\neq 0$, we get three energy bands. The energy dispersions around the Dirac points (${\bf K}$ or ${\bf K^\prime}$) for a particular value of $\alpha$ (intermediate to graphene and dice), namely, $\alpha=0.4$ are plotted in Fig. \ref{fig:soc_alphap4}.

\subsection{Results and discussion}
\subsubsection{No magnetic field}

The energies as a function of ring radius $R$ at the ${\bf K}$-valley is shown in Fig. \ref{fig:soc_alphap4}(a) and the same at the ${\bf K^\prime}$-valley is presented in Fig. \ref{fig:soc_alphap4}(b) with $m=0$ (green curves), $m=-1,-2$ (red curves) and $m=1,2$ (blue curves) for the valence and the conduction bands. Further, for distorted flat bands with $m=0$ (cyan curves), $m=-1,-2$ (black curves) and $m=1,2$ (yellow curves). It is seen that for $n=1$, the original dispersionless flat band is no longer flat, instead, it becomes distorted, i.e., a non-zero group velocity may be associated with the $n=1$ band electrons. This distortion implies that the $n = 1$ band electrons will start contributing to the transport properties of the system. Moreover, the $n=1$ band shifts away from the band center ($E=0$) and the bands for opposite spin species move in opposite directions (as indicated by dashed and solid lines in each panel), resulting in formation of spin polarized bands. However, it is worth noting that the spin polarizations for the ${\bf K}$- and ${\bf K^\prime}$-valleys are opposite. This certainly implies that the time-reversal symmetry remains unchanged. The spin-valley splitting of the $\alpha$-$T_3$ band structure bears resemblance to those observed in transition metal dichalcogenide materials, such as, MoS$_2$ [\onlinecite{spin_TMD}], where the two valleys exhibit opposite spin splitting while the system overall maintains a time-reversal symmetry. This intriguing spin-valley-dependent band structure can be experimentally observed through spin-valley-selective circular dichroism [\onlinecite{spin-Silicene}], a phenomenon in which the responses of the left and the right-handed circularly polarized light differ. For the graphene ($\alpha=0$) or the dice lattice ($\alpha=1$), the energy bands retain the spin and the valley degeneracy, that is, $E_{\zeta \sigma}=E_{\bar{\zeta} \bar{\sigma}}$ with $\bar{\sigma}=-\sigma$ and $\bar{\zeta}=-\zeta$ from Eq. (\ref{Energy_soc}), because the particle-hole symmetry remains invariant although the inversion symmetry is broken.

\begin{widetext}

\begin{figure}[h!]
\centering
\includegraphics[width=15cm, height=11cm]{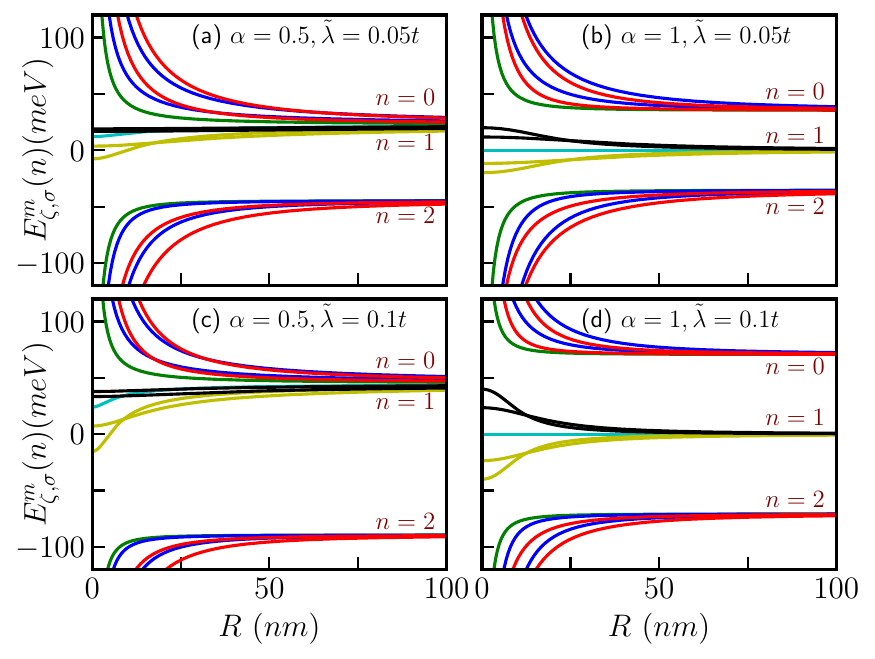}
\caption{(Color online) The energy dispersion at zero field as a function of $R$ at the ${\bf K}$-valley of the $\alpha$-$T_3$ quantum ring corresponding to the ISOC values (a) $\alpha=0.5$, $\tilde{\lambda}=0.05t$, (b) $\alpha=1$, $\tilde{\lambda}=0.05t$, (c) $\alpha=0.5$, $\tilde{\lambda}=0.1t$, and (d) $\alpha=1$, $\tilde{\lambda}=0.1t$ and $t=1eV$ are shown. The indices $n=0,1,$ and $2$ denote the CB, distorted FB, and VB respectively.}
\label{fig:soc_single}
\end{figure}
\end{widetext}

In Fig. \ref{fig:soc_single}, we present the energy bands at a particular valley (${\bf K}$-valley) and for $\uparrow$-spin only, by considering two different cases, namely, $\alpha=0.5$ and $\alpha=1$, corresponding to two values of the SOC term, namely, $\tilde{\lambda}=0.05t$ and $\tilde{\lambda}=0.1t$. The purpose is to observe how the band structure changes with $\alpha$ and responds to the SOC parameter $\tilde{\lambda}$. Let us first discuss the $\alpha=0$ (graphene) case [\onlinecite{Bol}]. In this scenario, the $n=1$ band (distorted flat band) shows no dispersion for all values of $m$, while all the branches of CB ($n=0$) and the VB ($n=2$) display a 1/R dependence in the small radius limit, whereas, $E_{\zeta,\sigma}\to \pm \tilde{\lambda}$ for large radii (not shown here). Notably, the criterion for large $R$ differs for different values of $m$. Additionally, the branches are two-fold degenerate with $m=(0,1), (2,-1), (3,-2)$ and so on. It is worth mentioning that all the energy levels are non-degenerate for all values of $\alpha$, except for $\alpha=0$. Further, at the ${\bf K}$-valley, as we increase $\alpha$, the distortion increases in the limit of small radius. However, in the large $R$ limit, some interesting features emerge as stated in the following. In the range $0<\alpha\leq 0.5$, the flat band ($n=1$) shifts away from the band center towards the CB ($n=0$), and this shift increases with $\alpha$. The $n=1$ bands merge to a value $\alpha \tilde{\lambda}$ and the $n=0$ band merge to $(1-\alpha)\tilde{\lambda}$ at the large $R$ limit. On the other hand, in the range $0.5<\alpha\leq 1$, the shift decreases with increasing $\alpha$. Again, the $n=1$ bands merge to a value $(1-\alpha)\tilde{\lambda}\cos\xi$ and the $n=0$ band merge at $\alpha\tilde{\lambda}\cos\xi$ in the large $R$ limit.  While the energy levels corresponding to $n=2$ merge at $-\tilde{\lambda}\cos\xi$ in the limit of large $R$, irrespective the value of $\alpha$. (For more details see Fig. \ref{fig:appa} of Appendix A). However, at the ${\bf K^\prime}$-valley the distorted flat bands shift towards the VB ($n=2$) and the bands corresponding to $n=0$ merge to a value $\tilde{\lambda}\cos\xi$ at the large $R$ limit (see solid curves in Fig. \ref{fig:soc_alphap4}(b)).

In Figs. \ref{fig:soc_single}(a),(c) we show the results for $\alpha=0.5$ with $\tilde{\lambda}=0.05t$ and $\tilde{\lambda}=0.1t$, respectively. Inside the distorted flat band ($n=1$), the degree of distortion increases with increasing $\tilde{\lambda}$ and the bands characterized by $m=0$ (cyan curves) and $m=1,2,..$ (yellow curves) exhibit large distortion. In contrast, the energy bands with  $m=-1,-2,..$(black curves) remain nearly dispersionless. In addition, we find the following interesting results for the dice lattice $\alpha=1$ (see Figs. \ref{fig:soc_single}(b),(d)). In the limit of small radii, all the branches of VB and CB vary as 1/R. Further, the branches are non-degenerate. Inside the $n=1$ flat band, we find the level with $m = 0$ remains flat at $E=0$. From Eq. (\ref{Energy_soc}), it may be noted that the other levels satisfy, $E_{\zeta,\sigma}^m(1)=-E_{\zeta,\sigma}^{-m}(1)$. Further, we have $E_{\zeta,\sigma}^m(0) = -E_{\zeta,\sigma}^{-m}(2)$ which are in direct contrast to the results for the bare case ($\tilde{\lambda}=0$). Here, only the $\uparrow$-spin and the ${\bf K}$-valley band structures are plotted, moreover, others for the $\downarrow$-spin and the ${\bf K^\prime}$-valley are essentially the same.
\\

\subsubsection{In presence of magnetic field}

Persistent equilibrium currents provide valuable information about the energy spectrum of a system near the Fermi energy. Although such currents are typically small and are detected through the magnetic moment they generate [\onlinecite{Bles}], recent experiments have shown promising results by employing dense arrays of rings on a cantilever, which enhances the magnetic signal and allows for both current measurements and the use of the setup as a sensitive magnetometer. The Corbino disk geometry, which can be fabricated with high precision using innovative techniques [\onlinecite{Ci2}] involving nanoparticle manipulation and hydrogenation of open bonds, is particularly suitable for studying persistent currents. To study the persistent currents and the effects of the magnetic field on the energy spectra, we include a magnetic flux threading the ring in the transverse direction, (${\bf B}=B_0\hat{z}$, where $B_0$ is a constant). This is incorporated in the Hamiltonian in Eq. (\ref{Eq.isoc_Ham}) as,
\begin{widetext}
\begin{equation}
H_{ring}^\Phi(\sigma)=\begin{pmatrix}
-\zeta\sigma\tilde{\lambda} \cos\xi & -i\frac{\hbar v_F}{R} (m-\zeta/2+\Phi/\Phi_0) \cos\xi & 0\\
i\frac{\hbar v_F}{R} (m-\zeta/2+\Phi/\Phi_0)\cos\xi & \zeta\sigma\tilde{\lambda}\cos\xi - \zeta\sigma\tilde{\lambda}\sin\xi & -i\frac{\hbar v_F}{R} (m+\zeta/2+\Phi/\Phi_0)\sin\xi\\
0 & i\frac{\hbar v_F}{R} (m+\zeta/2+\Phi/\Phi_0) \sin\xi & \zeta\sigma \tilde{\lambda} \sin\xi\\
\end{pmatrix}
\end{equation}
\end{widetext}
where $\Phi=\pi R^2B_0$, is the magnetic flux, and $\Phi_0$ is the usual flux quantum. Owing to this field, the modifications to the spectra can be written as,
\begin{equation}
\label{Energy_flux}
E_{\zeta,\sigma}^{m,\Phi}(n) = 2\sqrt{-\frac{P_\Phi}{3}}\cos\left[\frac{1}{3}\cos^{-1}\left(\frac{3Q_\Phi}{2P_\Phi}\sqrt{-\frac{3}{P_\Phi}} \right)-\frac{2\pi n}{3} \right]
\end{equation}
where $n$ = 0, 1, and 2 represent for conduction band (CB), flat band (FB), and valence band (VB) respectively. Here, the quantities $P$ and $Q$ are denoted by, 
\begin{align*}
P_\Phi=\frac{\tilde{\lambda}^2}{2}\sin2\xi-\tilde{\lambda}^2-\frac{\hbar^2v_F^2}{R^2}\bigg[N_\Phi^2\sin^2\xi+M_\Phi^2\cos^2\xi\bigg]
\end{align*}
and
\begin{multline*}
Q_\Phi=\frac{\zeta\sigma\tilde{\lambda}}{2}\sin2\xi\bigg[\tilde{\lambda}^2(\cos\xi-\sin\xi)+\frac{\hbar^2v_F^2}{R^2}\\
\big(M_\Phi^2\cos\xi-N_\Phi^2\sin\xi\big) \bigg]
\end{multline*}
with $M_\Phi=(m+\Phi/\Phi_0-\zeta/2)$, $N_\Phi=(m+\Phi/\Phi_0-\zeta/2)$. Thus the quantities $M_\Phi$ and $N_\Phi$ include $\Phi$, namely the flux.

The normalized spinor wavefunctions for the up and down spin are given by,

\begin{widetext}

\begin{equation*}
\Psi^{m\zeta\Phi}_{n,\sigma}(R,\theta)=N_{n,\sigma}^{\zeta\Phi}\, e^{im\theta}\begin{pmatrix}
-i\frac{\hbar v_F}{R} M_\Phi(E_{\zeta,\sigma}^{m,\Phi}-\zeta\sigma\tilde{\lambda}\sin\xi) \cos\xi\, e^{-i\zeta\theta}\\
(E_{\zeta,\sigma}^{m,\Phi}+\zeta\sigma\tilde{\lambda}\cos\xi)(E_{\zeta,\sigma}^{m,\Phi}-\zeta\sigma\tilde{\lambda}\sin\xi)\\
i\frac{\hbar v_F}{R} N_\Phi(E_{\zeta,\sigma}^{m,\Phi}+\zeta\sigma\tilde{\lambda}\cos\xi) \sin\xi\, e^{i\zeta\theta}
\end{pmatrix},
\end{equation*}
with,
\begin{equation}
\label{WaveFn_flux}
N_{n,\sigma}^{\zeta\Phi}=\frac{1}{\sqrt{\frac{\hbar^2v_F^2}{R^2}\Big[M_\Phi^2(E_{\zeta,\sigma}^{m,\Phi}-\zeta\sigma\tilde{\lambda}\sin\xi)^2 \cos^2\xi+N_\Phi^2(E_{\zeta,\sigma}^{m,\Phi}+\zeta\sigma\tilde{\lambda}\cos\xi)^2 \sin^2\xi\Big]+(E_{\zeta,\sigma}^{m,\Phi}+\zeta\sigma\tilde{\lambda}\cos\xi)^2(E_{\zeta,\sigma}^{m,\Phi}-\zeta\sigma\tilde{\lambda}\sin\xi)^2}}.
\end{equation}
\end{widetext}

\begin{figure}[h!]
\centering
  \includegraphics[width=\linewidth, height=\linewidth]{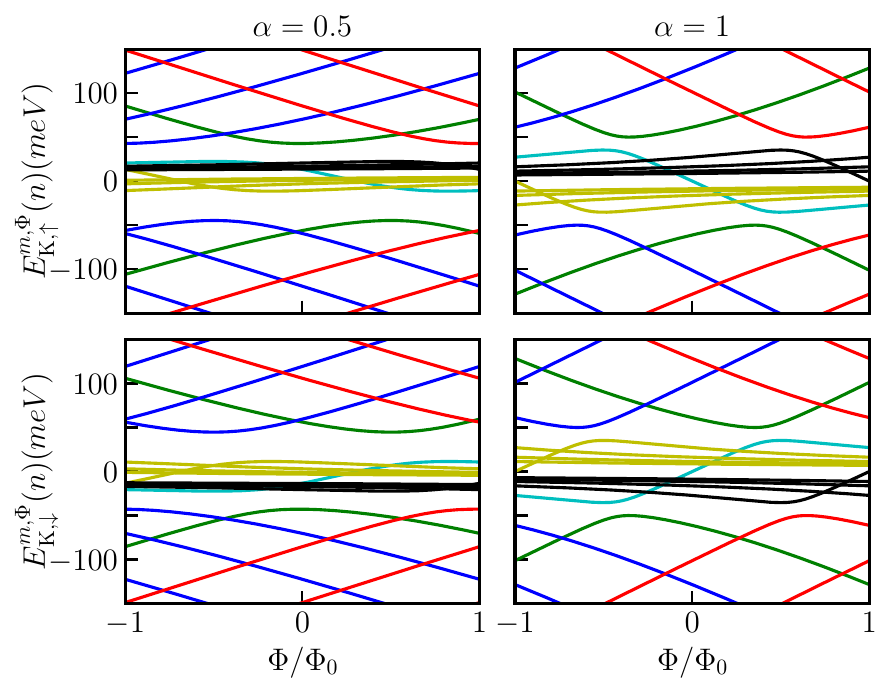}
  \caption{(Color online) The energy $E_{{\bf K},\sigma}^{m,\Phi}(n)$ as a function of external magnetic flux at ${\bf K}$-valley for different values of $\alpha$ with $R=10$ $nm$. Plots in the upper panel are for $\uparrow$-spin and the lower panel are for $\downarrow$-spin bands.}
  \label{fig:EvsPhi_spin}
\end{figure}  
  \begin{figure}[h!]
  \includegraphics[width=\linewidth, height=\linewidth]{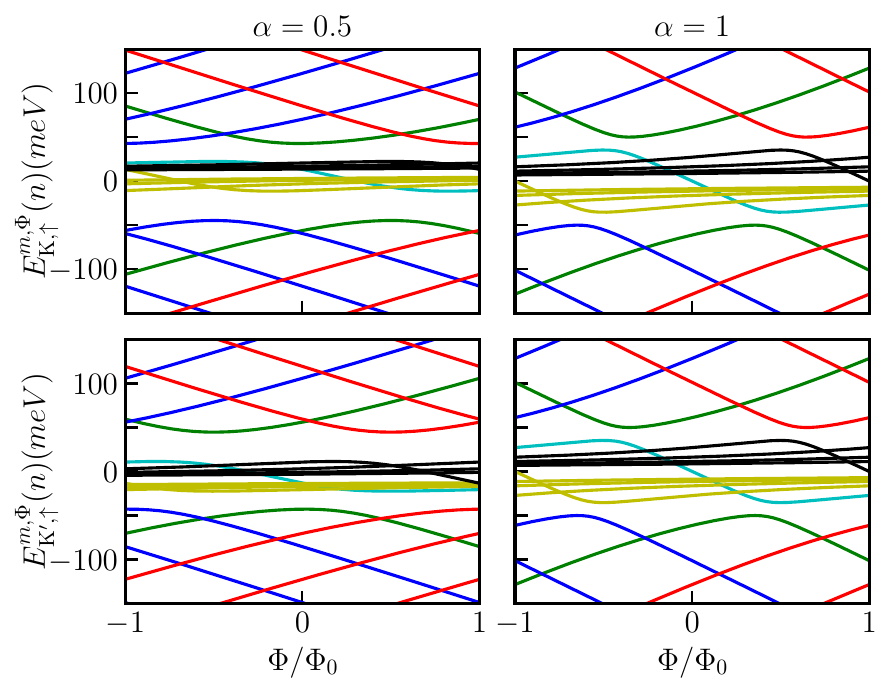}
  \caption{(Color online) The energy $E_{\zeta,\uparrow}^{m,\Phi}(n)$ as a function of external magnetic flux for $\uparrow$-spin bands for different values of $\alpha$ with $R=10$ $nm$. Plots in the upper panel are for ${\bf K}$-valley and the lower panel are for ${\bf K^\prime}$-valley.}
  \label{fig:EvsPhi_valley}
\end{figure}


The qualitative features of the energy spectrum deviate significantly from the case corresponding to zero magnetic field. In contrast to the $B_0 = 0$ situation, we observe that all energy levels are non-degenerate for all values of $\alpha$. Additionally, we find that $E_{\zeta,\sigma}^{m,\Phi}(1)\neq-E_{\zeta,\sigma}^{-m,\Phi}(1)$, $E_{\zeta,\sigma}^{m,\Phi}(0) = -E_{\zeta,\sigma}^{-m,\Phi}(2)$. Furthermore, there is distortion of the $m = 0$ flat band energy level for $\alpha = 1$. In addition, the flat band energy levels merge with zero energy at large values of $R$. However, this is not the scenario for intermediate values of $\alpha$. As $\alpha \neq 0$, the zero energy modes merge at some finite (positive) energy values depending upon the value of $\alpha$ in the large $R$ limit. Again each of the levels of the VB and CB exhibits a non-monotonic behaviour as a function of the radius $R$. The energy levels attain an extremum for a given value of $m$ (minimum for the conduction band and maximum for the valence band) at a particular value of $R$. However, positions of these extrema depend on the value of $m$. For a particular $m$, in the limit of small $R$, all the energy levels vary inversely as $R$. On the other hand, the energy scales as, $E \sim R$ in the limit of large $R$. More detailed results are provided in the Fig. \ref{fig:appb} of Appendix B.

The energy levels as a function of the external magnetic flux are shown in Fig. \ref{fig:EvsPhi_spin} and Fig. \ref{fig:EvsPhi_valley} for a quantum ring with a particular radius and SOC, namely, $R=10$ $nm$ and $\tilde{\lambda}=0.05t$. We consider two cases, for a particular valley (say, ${\bf K}$-valley) with both $\uparrow$- and $\downarrow$- spins, and for a particular spin (say $\uparrow$-spin) involving both the ${\bf K}$- and the ${\bf K^\prime}$-valleys. It can be observed that for $\alpha=0$ at ${\bf K}$-valley, $E_{{\bf K},\uparrow}^{m,\Phi}(n)=E_{{\bf K},\downarrow}^{m,\Phi}(n)$ (not shown here), thus it will possess zero spin current (we shall discuss it later). From Fig. \ref{fig:EvsPhi_spin}, it is obsered that for non-zero $\alpha$, we get, $E_{{\bf K},\uparrow}^{m,\Phi}(0)=-E_{{\bf K},\downarrow}^{m,\Phi}(2)$, $E_{{\bf K},\uparrow}^{m,\Phi}(2)=-E_{{\bf K},\downarrow}^{m,\Phi}(0)$ and for the distorted flat band, we have $E_{{\bf K},\uparrow}^{m,\Phi}(1)=-E_{{\bf K},\downarrow}^{m,\Phi}(1)$. Moreover, for any intermediate $\alpha$ ($0<\alpha<1$) in the case of $\uparrow$-spin, the $n=1$ bands are shifted towards $n=0$ CBs, while the $\downarrow$-spin bands are shifted towards the $n=2$ VBs, as discussed earlier. For the dice case ($\alpha=1$), the $n=1$ bands are equally distributed towards both the CB and VB.

 Moving on to Fig. \ref{fig:EvsPhi_valley}, we observe following the interesting aspects. For any value of $\alpha$ such that $\alpha<1$, $E_{\zeta,\uparrow}^{m,\Phi}(n) \neq E_{-\zeta,\uparrow}^{m,\Phi}(n)$ since the effective time reversal symmetry (TRS) is broken by the external magnetic field. However, for the dice lattice we can see $E_{\zeta,\uparrow}^{m,\Phi}(n) = E_{-\zeta,\uparrow}^{m,\Phi}(n)$ for all the bands. Also, for any intermediate $\alpha$ ($0<\alpha<1$) the $\uparrow$-spin bands of the $n=1$ bands are shifted towards $n=0$ bands (CBs), while the $\downarrow$-spin bands are shifted towards the $n=2$ bands (VBs) in the ${\bf K^\prime}$-valley as discussed earlier. Again for the dice case ($\alpha=1$), the $n=1$ bands are equally spread towards the CBs and VBs for both the valleys.


\subsubsection{Charge persistent currents}
\begin{figure}[h!]
\centering
  \includegraphics[width=\linewidth, height=\linewidth]{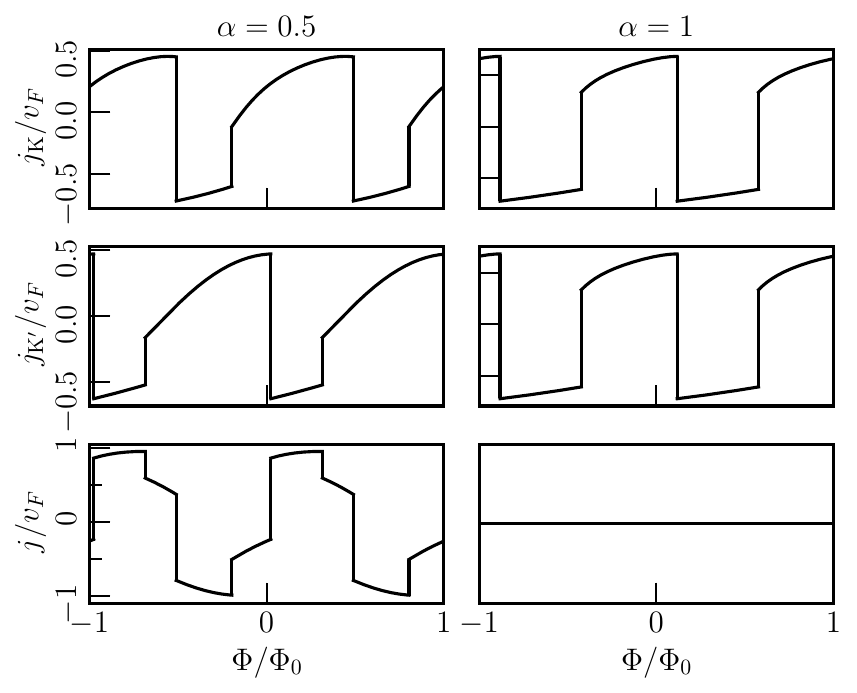}
  \caption{(Color online) The charge persistent currents as a function of external magnetic flux $\Phi/\Phi_0$ for $\uparrow$-spin bands are shown. Upper row is for the $\mathrm{K}$-valley, middle row is for the $\mathrm{K^\prime}$-valley, and the lower row shows the valley current. Here we have considered $\tilde{\lambda}=0.05t$, $t=1eV$, and $R=10$ $nm$. We have shown results for two different $\alpha$ values, namely, $\alpha=0.5$ and $1$.}
  \label{fig:valley_cur}
\end{figure}
Let us delve into the topic of persistent currents. The persistent current is the equilibrium current that flows along the angular ($\theta$) direction in a QR when threaded by a magnetic flux. This current can be calculated using the relation, $j_{x(y)} = v_F [\Psi^\dagger S_{x(y)}\Psi]$, where $S_{x,y}$ denote the $x$ and $y$ components of the pseudospin operator. Using this definition, the radial and the angular currents are further obtained as $j_r = v_F [\Psi^\dagger S_r \Psi]$ and $j_\theta = v_F [\Psi^\dagger S_\theta \Psi]$, respectively. Here, $S_r$ and $S_\theta$ are given by $S_r = S_x \cos\theta + S_y \sin \theta$ and $S_\theta = -S_x\sin\theta + S_y \cos\theta$, respectively. Along a particular direction ($r = R$) of the $\alpha$-$T_3$ ring, although the radial current vanishes, we can calculate the angular current at a particular valley as well as for a particular spin. The analytical expression for the charge persistent current is derived as follows,
\begin{widetext}

\begin{equation}
\label{P_cur}
\resizebox{\hsize}{!}{$
j_n^\zeta=2v_F\frac{\hbar v_F}{R}{N_{\kappa,\sigma}^{\zeta\Phi}}^2\big(E_{\zeta,\sigma}^{m,\Phi}(n)+\zeta\sigma\tilde{\lambda}\cos\xi\big)\big(E_{\zeta,\sigma}^{m,\Phi}(n)-\zeta\sigma\tilde{\lambda}\sin\xi)\Big[M_\Phi\big(E_{\zeta,\sigma}^{m,\Phi}(n)-\zeta\sigma\tilde{\lambda}\sin\xi\big)\cos^2\xi+N_\Phi\big(E_{\zeta,\sigma}^{m,\Phi}(n)+\zeta\sigma\tilde{\lambda}\cos\xi\big)\sin^2\xi\Big]$}
\end{equation}

\end{widetext}

The net persistent current at a particular valley is calculated by considering the contributions from the valence band ($n = 2$) and the distorted flat band ($n = 1$) via, $j_\zeta =j_{n=2}^\zeta + j_{n=1}^\zeta $. It is worthy to mention that the distortion of the energy levels in the flat band gives rise to a finite persistent current. From Fig. \ref{fig:EvsPhi_valley} it is observed that the low-energy state comprises of different $m$ values for a range of the scaled flux, $\Phi/\Phi_0$. Using this low-energy state from Fig. \ref{fig:EvsPhi_valley}, we illustrate the variation of the persistent currents as a function of $\Phi/\Phi_0$ with $\tilde{\lambda} = 0.05t$ and $R = 10$ $nm$ in Fig. \ref{fig:valley_cur}. The inclusion of the spin-orbit coupling term alters the oscillation pattern of the persistent current compared to bare ($\tilde{\lambda}=0$) case [\onlinecite{Mijanur}]. In the upper panel of Fig. \ref{fig:valley_cur} we display the persistent current due to the $\uparrow$-spin at the ${\bf K}$-valley, while in the middle panel, we show the same at the ${\bf K^\prime}$-valley. The currents at different valleys are no longer equal, except for the cases $\alpha=0$ (not shown here) and $\alpha=1$. Moreover, the persistent current at a specific valley oscillates periodically with $\Phi/\Phi_0$, with the periodicity $\Phi/\Phi_0 = 1$. Though the oscillation pattern varies with the parameter $\alpha$, the oscillation period remains independent of $\alpha$. Additionally, we observe a finite persistent current at a specific valley when $\Phi = 0$ (no magnetic field) for all values of $\alpha$.

We further introduce a quantity called the valley current as,
\begin{equation}
j=j_Q(\zeta=1)-j_Q(\zeta=-1)
\end{equation}
where $\zeta$ denotes the valley index. In this context, we find that the persistent current corresponding to the two extreme cases ($\alpha = 0$ (not shown here) and $\alpha = 1$) are equal, resulting in zero valley current for graphene and the dice lattice. However, for an intermediate value of $\alpha$ (say, $\alpha=0.5$), we observe an oscillatory valley current with the oscillation period of $\Phi/\Phi_0=1$. Furthermore, the oscillation pattern of the valley current depends on the parameter $\alpha$.

\subsubsection{Equilibrium spin currents}
\begin{figure}[h!]
\includegraphics[width=\linewidth, height=\linewidth]{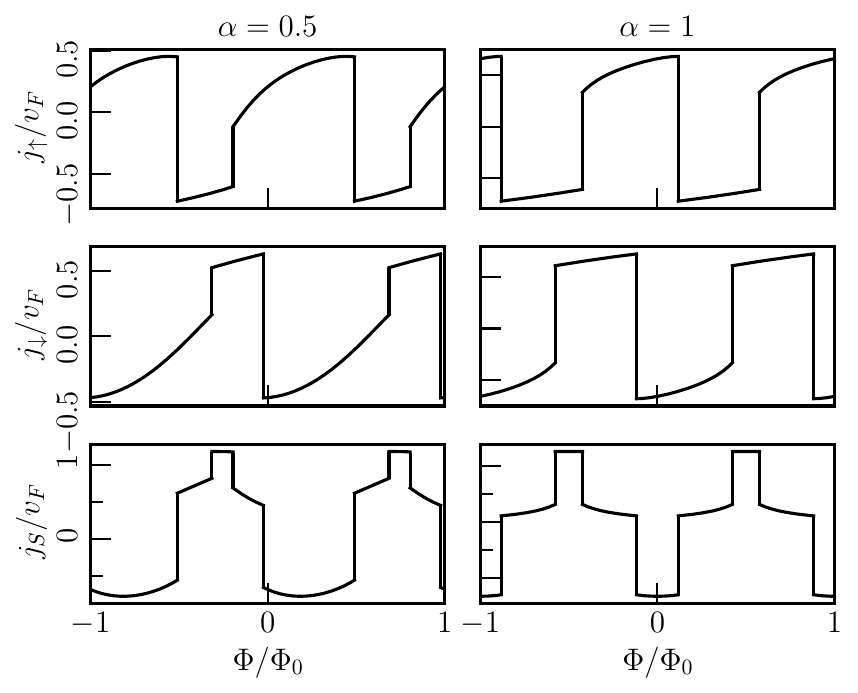}
  \caption{(Color online) The equilibrium spin current as a function of $\Phi/\Phi_0$ are shown. Plots in the upper row are for the $\uparrow$-spin, middle row for the $\downarrow$-spin of ${\bf K}$-valley, and the bottom row contain the spin current at the same valley. Here we have considered $\tilde{\lambda}= 0.05t$, $t=1eV$, and $R=10$ $nm$. We have shown for two different $\alpha$ values, namely, $\alpha=0.5$ and $1$.}
  \label{fig:spin_cur}
\end{figure}

Now, we turn our attention to calculation of the equilibrium spin currents. We define the equilibrium spin current as,
\begin{equation}
j_S=j_n^{\zeta}(\sigma=1)-j_n^{\zeta}(\sigma=-1)
\end{equation}
where $\sigma=\pm 1$ denotes $\uparrow$- and $\downarrow$-spins. As mentioned earlier, for non-zero $\alpha$, the spin resolved energy levels are not the same within a range of flux in unit of $\Phi/\Phi_0$, thus potentially leading to formation of spin currents. To investigate this, we first calculate the persistent current for both the $\uparrow$-spin and $\downarrow$-spin bands at a particular valley, using the same procedure as mentioned above. In the upper panel of Fig. \ref{fig:spin_cur}, we show the persistent current for $\uparrow$-spin, while in the middle panel, we show the same for $\downarrow$-spin at the ${\bf K}$-valley. We observe that both the $\uparrow$-spin and $\downarrow$-spin persistent currents oscillate with $\Phi/\Phi_0$ with a periodicity of $\Phi/\Phi_0=1$ for all values of $\alpha$. Thus, this result is distinct from the valley current. Additionally, the currents corresponding to different spins are no longer equal, except when $\alpha=0$. The latter yields a zero spin current ($j_S$) for $\alpha=0$ (not shown here). Further, for non-zero $\alpha$, we observe oscillatory spin current with the oscillation period of $\Phi/\Phi_0=1$. However, the oscillation pattern of the spin current is affected by the value of $\alpha$, as shown in the lower panel of Fig. \ref{fig:spin_cur}.

\section{Pseudospin-1 QR with RSOC}
\label{Rash}

Here we consider the system to have Rashba SOC. The corresponding Hamiltonian can be written as, $H=H_0+H_R$, where $H_0$ is the tight-binding term, and $H_R$ is the Rashba spin-orbit coupling term. We write the Hamiltonian as,
\begin{widetext}
\begin{equation}
\label{Ham_Rash}
H=-\sum_{\langle ij\rangle\sigma}tc_{i\sigma}^\dagger c_{j\sigma}-\sum_{\langle ik\rangle\sigma}\alpha t c_{i\sigma}^\dagger c_{k\sigma}-\sum_{\langle ij\rangle\sigma\sigma^\prime}i\lambda_Rc_{i\sigma}^\dagger(\hat{D}_{ij}\cdot\vec{\tau})_{\sigma\sigma^\prime}c_{j\sigma^\prime}-\sum_{\langle ik\rangle\sigma\sigma^\prime}i\alpha\lambda_Rc_{i\sigma}^\dagger(\hat{D}_{ik}\cdot\vec{\tau})_{\sigma\sigma^\prime}c_{k\sigma^\prime}+\bigg (h.c.\bigg)
\end{equation}
\end{widetext}
where $\sigma = \uparrow, \downarrow$, spin indices and $i,j,k$ are labels for sites corresponding to A, B, and C sublattices respectively. The first term is the electron hopping between the A and B sites, while the second one is that between the A and C sites. The summation of $\langle ij \rangle$ ($\langle ik \rangle$) runs over the nearest neighbour sites of AB (AC). Further, the Rashba SOC induced by electric fields due to a gradient of the crystal potential [\onlinecite{Wa,Rahul1,Rahul2}].
Where $\vec{\tau}=\tau_x\hat{x}+\tau_y\hat{y}+\tau_z\hat{z}$ is the Pauli matrix vector, $\hat{D}_{ij}$ ($\hat{D}_{ik}$) is the unit vector along the direction of the cross product $\vec{E}_{ij} \times \vec{r}_{ij}$ ($\vec{E}_{ik} \times \vec{r}_{ik}$) of the electric field $\vec{E}_{ij}$ ($\vec{E}_{ik}$) and displacement $\vec{r}_{ij}$ ($\vec{r}_{ik}$) for the bond $ij$ ($ik$). $\lambda_R$ is the strength of Rashba SOC between the A and the B sites while $\alpha\lambda_R$ is that between the A and the C sites. Here we ignore the complex NNN hopping. In momentum space, the Hamiltonian of the $\alpha$-$T_3$ lattice becomes,
\begin{widetext}
\begin{equation}
H=\begin{pmatrix}
0 & -t\gamma_\mathrm{k}^* & 0 & 0 & -i\lambda_R\gamma_{\mathrm{k}+}^* & 0\\
-t\gamma_\mathrm{k} & 0 & -\alpha t\gamma_\mathrm{k}^* & i\lambda_R\gamma_{\mathrm{k}-} & 0 & i\alpha\lambda_R\gamma_{\mathrm{k}+}^*\\
0 & -\alpha t\gamma_\mathrm{k} & 0 & 0 & -i\alpha\lambda_R\gamma_{\mathrm{k}-} & 0\\
0 & -i\lambda_R\gamma_{\mathrm{k}-}^* & 0 & 0 & -t\gamma_\mathrm{k}^* & 0\\
i\lambda_R\gamma_{\mathrm{k}+} & 0 & i\alpha\lambda_R\gamma_{\mathrm{k}-}^* & -t\gamma_\mathrm{k} & 0 & -\alpha t\gamma_\mathrm{k}^*\\
0 & -i\alpha\lambda_R\gamma_{\mathrm{k}+} & 0 & 0 & -\alpha t\gamma_\mathrm{k} & 0
\end{pmatrix}
\end{equation}
\end{widetext}
we defined $\gamma_\mathrm{k}=1+e^{ik_1}+e^{ik_2}$ and $\gamma_{\mathrm{k}\pm}=1+e^{i(k_1\pm 2\pi/3)}+e^{i(k_2\pm 4\pi/3)}$, where the components are along the axes indicated in Fig. \ref{fig:ring_geo_soi}(b) as $k_i=\vec{k}.\hat{\mathrm{e}}_i$. The annihilation operator basis is $(c_{1\mathrm{k}\uparrow},c_{2\mathrm{k}\uparrow},c_{3\mathrm{k}\uparrow},c_{1\mathrm{k}\downarrow},c_{2\mathrm{k}\downarrow},c_{3\mathrm{k}\downarrow})$. In the vicinity of a Dirac point (namely, {\bf K}), the Hamiltonian corresponding to an ideal $\alpha$-$T_3$ ring is given by  [\onlinecite{FE,Spin_dep1,Spin_dep2,Bol,DR,Graph_Model1, Graph_Model2,Mijanur}],
\begin{widetext}
\begin{equation}
\resizebox{\hsize}{!}{$
H_{ring}=\frac{\hbar v_F}{R}\begin{pmatrix}
0 & -i(m+\frac{1}{2})\cos\xi e^{\frac{i\pi}{3}} & 0 & 0 & -\frac{\lambda_R}{t}(m+\frac{1}{2})\cos\xi e^{\frac{i\pi}{3}} & 0\\
i(m+\frac{1}{2})\cos\xi e^{-\frac{i\pi}{3}} & 0 & -i(m-\frac{1}{2})\sin\xi e^{\frac{i\pi}{3}} & \frac{\lambda_R}{t}(m-\frac{1}{2})\cos\xi e^{\frac{i\pi}{3}} & 0 & -\frac{\lambda_R}{t}(m+\frac{1}{2})\sin\xi e^{-\frac{i\pi}{3}}\\
0 & i(m-\frac{1}{2})\sin\xi e^{-\frac{i\pi}{3}} & 0 & 0 & -\frac{\lambda_R}{t}(m-\frac{1}{2})\sin\xi e^{-\frac{i\pi}{3}} & 0\\
0 & \frac{\lambda_R}{t}(m-\frac{1}{2})\cos\xi e^{-\frac{i\pi}{3}} & 0 & 0 & i(m-\frac{1}{2})\cos\xi e^{-\frac{i\pi}{3}} & 0\\
-\frac{\lambda_R}{t}(m+\frac{1}{2})\cos\xi e^{-\frac{i\pi}{3}} & 0 & \frac{\lambda_R}{t}(m-\frac{1}{2})\sin\xi e^{\frac{i\pi}{3}} & -i(m-\frac{1}{2})\cos\xi e^{\frac{i\pi}{3}} & 0 & i(m+\frac{1}{2})\sin\phi e^{-\frac{i\pi}{3}}\\
0 & -\frac{\lambda_R}{t}(m+\frac{1}{2})\sin\xi e^{\frac{i\pi}{3}} & 0 & 0 & -i(m+\frac{1}{2})\sin\xi e^{\frac{i\pi}{3}} & 0
\end{pmatrix}$}
\end{equation}
\end{widetext}
where $\tan\xi=\alpha$ and $\hbar v_F=3at/2\cos\xi$. The eigenstates of the ring Hamiltonian can be obtained as,
\begin{equation}
\psi(R,\theta)=\begin{pmatrix}
\chi_{1\uparrow}(R)e^{i(m+1)\theta}\\
\chi_{2\uparrow}(R)e^{im\theta}\\
\chi_{3\uparrow}(R)e^{i(m-1)\theta}\\
\chi_{1\downarrow}(R)e^{i(m-1)\theta}\\
\chi_{2\downarrow}(R)e^{im\theta}\\
\chi_{3\downarrow}(R)e^{i(m+1)\theta}
\end{pmatrix}
\end{equation}
where the integer $m$ labels the orbital angular momentum quantum number and $\chi_i$ denotes the amplitudes corresponding to the three sublattices. Here, we investigate the behaviour at a given value of radius $r$, namely $r = R$, such that the radial part is rendered frozen in the eigensolution. Again for the sake of the hermiticity of the Hamiltonian in ring geometry we made the replacements $r \to R$ and $\frac{\partial}{\partial r} \to -\frac{1}{2R}$. We further obtain the energy spectrum as,
\begin{equation}
\label{Egn_Rash}
\begin{aligned}
E_{1} &= 0\\
E_{2} &= \kappa \frac{\epsilon}{2}\sqrt{(1+4m^2-4m\frac{1-\alpha^2}{1+\alpha^2})(1+\frac{\lambda_R^2}{t^2})}\\
E_{3} &= \kappa \frac{\epsilon}{2}\sqrt{(1+4m^2)(1+\frac{\lambda_R^2}{t^2}\frac{1-\alpha^2}{1+\alpha^2})+4m(\frac{\lambda_R^2}{t^2}+\frac{1-\alpha^2}{1+\alpha^2})}
\end{aligned}
\end{equation}
where $\kappa=\pm 1$ is the particle-hole index and $\epsilon=\frac{\hbar v_F}{R}$. $E_1$ is the zero energy flat band, $E_2$ is the $\uparrow$-spin energy band and $E_3$ is the $\downarrow$-spin band.
\begin{widetext}

\begin{figure}[h!]
\centering
\includegraphics[width=15cm, height=10.8cm]{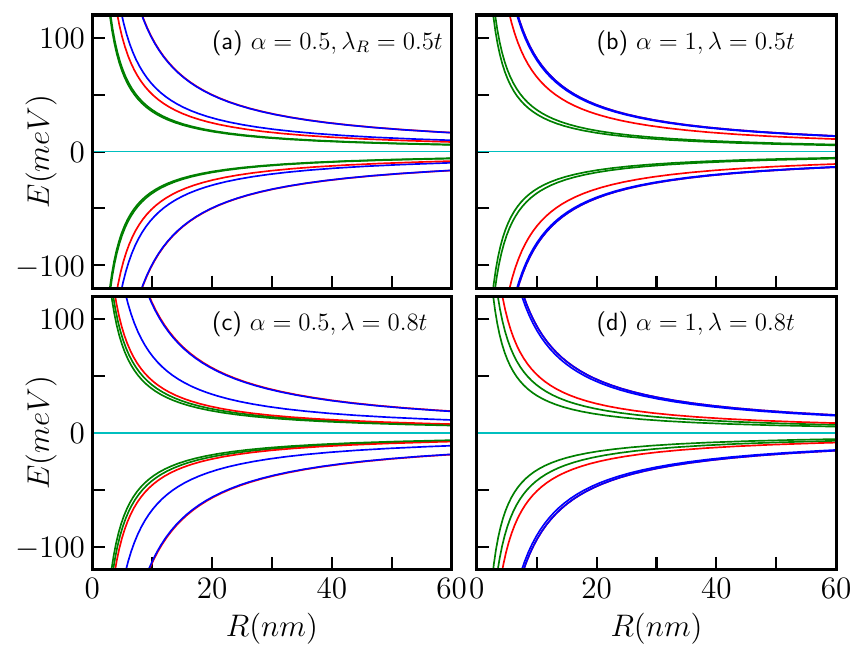}
\caption{(Color online) The energy levels in the absence of an external field with $m=-1$ (red curves), $0$ (green curves), and $1$ (blue curves) of the $\alpha$-$T_3$ quantum ring as a function of ring radius $R$ in presence of only RSOC are shown. (a) for $\alpha=0.5$, $\lambda_R=0.5t$, (b) for $\alpha=1$, $\lambda_R=0.5t$, (c) for $\alpha=0.5$, $\lambda_R=0.8t$, and (d) for $\alpha=1$, $\lambda_R=0.8t$. We have taken $t=1eV$.}
\label{fig:Eng_Rash}
\end{figure}
\end{widetext}
\subsection{Results and discussions}

\subsubsection{No magnetic field}
The energy spectra at the {\bf K}-valley in presence of Rashba SOC are expressed in Eq. (\ref{Egn_Rash}). Fig. \ref{fig:Eng_Rash} displays the energies as a function of the ring radius, R, for various values of $\alpha$ and $\lambda_R$. One can easily verify the results of the $\alpha$-$T_3$ quantum ring without the RSOC term by setting $\lambda_R = 0$ in Eq. (\ref{Egn_Rash}). In this case, we have considered two scenarios, namely, $\lambda_R = 0.5t$ and $\lambda_R=0.8t$ corresponding to $\alpha=0.5$ and $1$. We have plotted only the $m = -1$, $0$, and $1$ bands represented by red, green, and blue curves, respectively. When $\lambda_R=0$, the system exhibits three bands, with one being totally flat band. However, with a non-zero $\lambda_R$ the original three bands split into six spin-split bands, including two non-dispersive flat bands and four dispersive bands as described by Eq. (\ref{Egn_Rash}). From Fig. \ref{fig:Eng_Rash}, it is evident that all the energy branches have a 1/$R$ dependence and approach $E\to 0$ for very large radii, irrespective of the value of $\alpha$. Additionally, the dispersive bands remain non-degenerate, in contrast to the case of the pseudospin-1 $\alpha$-$T_3$ QR without SOC. Moreover, the dispersive $\uparrow$-spin and $\downarrow$-spin bands split as well. Specifically, for $m=0$, the energies are given by,
\begin{equation}
E_2=\frac{\kappa \epsilon}{2}\sqrt{1+\frac{\lambda_R^2}{t^2}}
\end{equation}
and 
\begin{equation}
E_{3}=\frac{\kappa\epsilon}{2}\sqrt{1+\frac{\lambda_R^2}{t^2}\frac{1-\alpha^2}{1+\alpha^2}}.
\end{equation}
It can be observed that the $\uparrow$-spin energy band, $E_{2}$ is independent of $\alpha$, while  the $\downarrow$-spin band, $E_{3}$ has a dependency on $\alpha$. Consequently, the splitting between the $m=0$ bands (green curves in Fig. \ref{fig:Eng_Rash}) increases with increasing values of $\alpha$. Whereas, the splitting between the bands with $m=-1$ (red curves in Fig. \ref{fig:Eng_Rash}) and $m=1$ (blue curves in Fig. \ref{fig:Eng_Rash}) decreases as $\alpha$ increases. Furthermore, the energy splitting decreases with the increase of $|m|$ values for all values of $\alpha$. In addition to that, the energy splitting increases with $\lambda_R$ increasing. An intriguing observation is that for $\alpha=1$, i.e., the dice lattice, the energies are
\begin{equation}
E_{2}=\frac{\kappa\epsilon}{2}\sqrt{(1+4m^2)(1+\frac{\lambda_R^2}{t^2})}
\end{equation}
and
\begin{equation}
E_{3}=\frac{\kappa\epsilon}{2}\sqrt{1+4m^2+4m\frac{\lambda_R^2}{t^2}}.
\end{equation}
Thus, $E_{2}$ is a even function of $m$, making it two-fold degenerate with $m=\pm1,\pm2, \pm3,....$ etc. On the other hand, the $E_{3}$ band is an odd function of $m$, resulting in its non-degeneracy as illustrated in Figs. \ref{fig:Eng_Rash}(b) and \ref{fig:Eng_Rash}(d).

\begin{widetext}

\begin{figure}[h!]
\centering
\includegraphics[width=15cm, height=11cm]{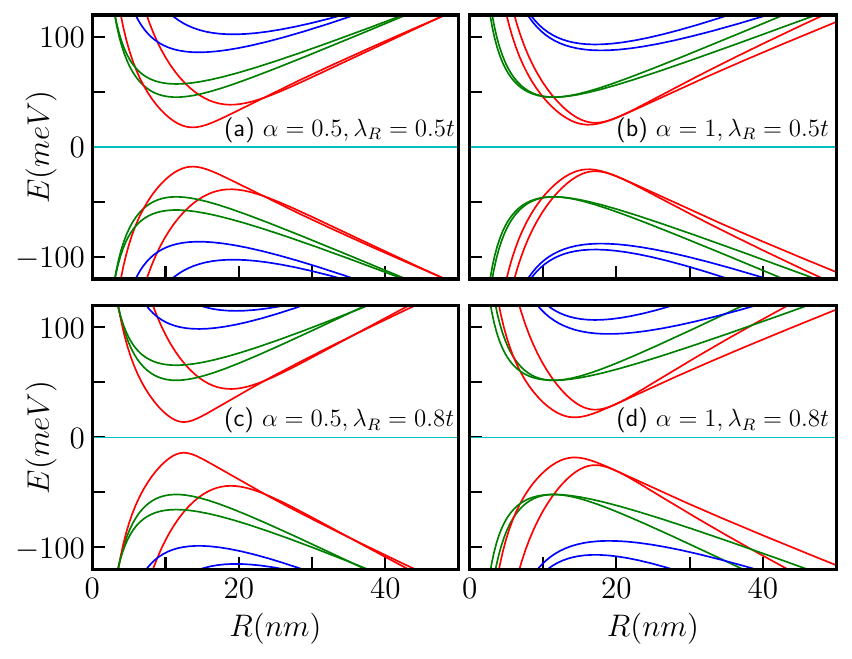}
\caption{(Color online) Energy levels with $m=-1$ (red curves), $0$ (green curves), and $1$ (blue curves) of the $\alpha$-$T_3$ quantum ring as a function of ring radius $R$ in presence of an external magnetic field of magnitude $B_0=5$T (a) for $\alpha=0.5$, $\lambda_R=0.5t$, (b) for $\alpha=1$, $\lambda_R=0.5t$, (c) for $\alpha=0.5$, $\lambda_R=0.8t$, and (d) for $\alpha=1$, $\lambda_R=0.8t$ and $t=1eV$ taken.}
\label{fig:Eng_B_Rash}
\end{figure}
\end{widetext}
\subsubsection{In presence of magnetic field}
Now let us discuss the case when the pseudospin-1 $\alpha$-$T_3$ ring is threaded by a perpendicular magnetic field ${\bf B}=B_0\hat{z}$. The spectrum of the system is modified by the field flux as follows,

\begin{equation}
\begin{aligned}
E_{1}(\Phi) &= 0\\
E_{2}(\Phi) &= \kappa\frac{\epsilon}{2}\biggl\{\bigg[1+4(m+\beta)^2-4(m+\beta)\frac{1-\alpha^2}{1+\alpha^2}\bigg]\\
& \bigg(1+\frac{\lambda_R^2}{t^2}\bigg)\biggl\}^\frac{1}{2}\\
E_{3}(\Phi) &= \kappa \frac{\epsilon}{2}\biggl\{\big[1+4(m+\beta)^2\big]\bigg(1+\frac{\lambda_R^2}{t^2}\frac{1-\alpha^2}{1+\alpha^2}\bigg)+\\
& 4(m+\beta)\bigg(\frac{\lambda_R^2}{t^2}+\frac{1-\alpha^2}{1+\alpha^2}\bigg)\biggl\}^\frac{1}{2}
\end{aligned}
\label{Egn_Rash_flux}
\end{equation}
where $\beta=\Phi/\Phi_0$ with $\Phi=\pi R^2B_0$ is magnetic flux through the ring and $\Phi_0$ is the usual flux quantum. The Zeeman coupling has been neglected at small enough values of the field. The addition of a magnetic field, represented by a U(1) minimal coupling with flux $\Phi$ threading the ring, breaks the time reversal symmetry allowing for the emergence of persistent charge currents [\onlinecite{Butt}] which we shall discuss later.

 In Fig. \ref{fig:Eng_B_Rash}, we show the dependence of a few energy levels on the ring radius, $R$, considering $B_0 = 5$T for the two aforementioned cases i.e., $\lambda_R = 0.5t$ with $\alpha=0.5,1$, and $\lambda_R=0.8t$ with $\alpha=0.5,1$. Each level exhibits a non-monotonic behaviour as a function of the radius $R$. The energy levels attain an extremum (minimum for conduction band and maximum for valence band) at a particular value of $R$. However, the positions of these extrema depend on the value of $m$, $\alpha$ and $\lambda_R$ explicitly. In the limit of small $R$, all the energy levels vary inversely with $R$. On the other hand, the energy scales as, $E \sim |R|$ that is in limit of large $R$. Additionally, for a fixed magnetic field and for large $m$, the extrema points depend on $R$ as $R\propto \sqrt{|m|}$ irrespective of $\alpha$. Thus, the concept of large radii differs for different values of $m$. Consequently, for negative values of $m$, the extrema points of the energy exhibit a scaling behaviour, namely, $E_{min}\propto 1/\sqrt{|m|}$, resulting in a diminishing of the spectral gap with increasing $|m|$. Conversely, for positive values of $m$, the energy extrema scales as, $E_{min}\propto \sqrt{m}$. Furthermore, from Eq. (\ref{Egn_Rash_flux}), it is evident that in presence of a magnetic field, the energy splitting between the bands of the $m = 0$ level as well as the $m \neq 0$ levels decreases with the increase in the values of the parameters $\alpha$ and $\lambda_R$. Again, from Eq. (\ref{Egn_Rash_flux}) it is noted that for $\alpha < 1$, there are two points where the spin bands cross each other as a function of $R$ for $m=0$ and $m=-1,-2,-3,...$ etc. bands, whereas there is only one band crossing point for $m=1,2,3...$ values. These crossings obey the following condition,
\begin{widetext}
\begin{equation}
\label{Eqn_flux}
\frac{\lambda_R^2}{t^2}\big[1+4(m+\beta)^2-4(m+\beta)\big]-\frac{\lambda_R^2}{t^2}\frac{1-\alpha^2}{1+\alpha^2}\big[1+4(m+\beta)^2+4(m+\beta)\big]-8(m+\beta)\frac{1-\alpha^2}{1+\alpha^2}=0
\end{equation}
\end{widetext}
Eq. (\ref{Eqn_flux}) can be checked against the plots shown in Figs. \ref{fig:Eng_B_Rash}(a) and \ref{fig:Eng_B_Rash}(c). Now, for the dice lattice case ($\alpha=1$), the above mentioned condition requires $m+\beta=1/2$, which implies that along the radius $R$, the band crossing point occurs at $R=\sqrt{2}l_0\sqrt{(1/2-m)}$, where $l_0=\sqrt{\hbar/(eB_0)}$ is the magnetic length. Consequently, there is only one band crossing point for $m=0$ and $m=-1,-2,-3,..$ etc. bands. Furthermore, band crossing is prohibited for $m$ to be positive as there are no real values of $R$ for $m>0$, indicating that the corresponding spin bands do not cross each other as illustrated in Figs. \ref{fig:Eng_B_Rash}(b) and \ref{fig:Eng_B_Rash}(d) by the blue curves. 

\begin{widetext}

\begin{figure}[h!]
\centering
\includegraphics[width=16cm, height=11.5cm]{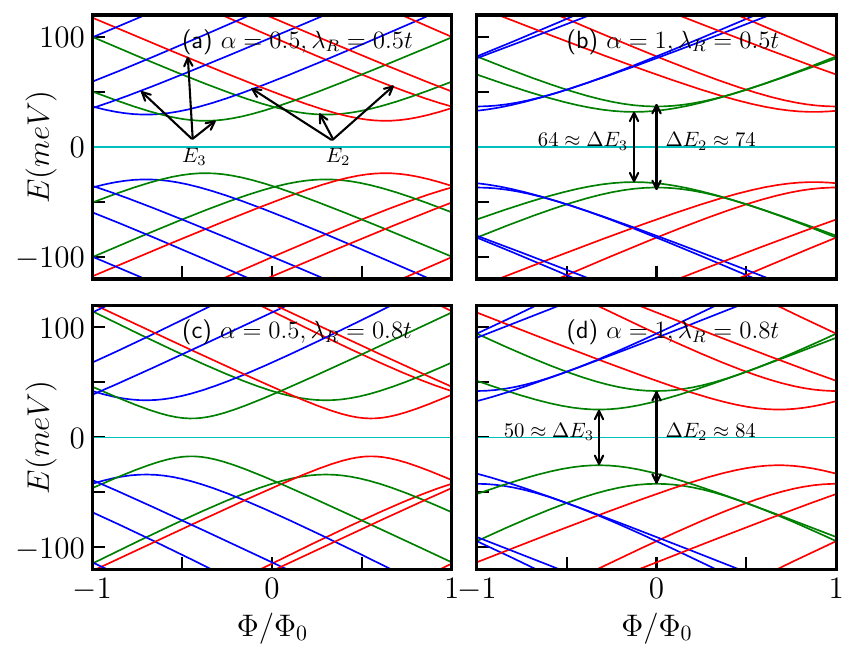}
\caption{(Color online) Energy levels as function of external magnetic flux $\Phi/\Phi_0$ for (a) for $\alpha=0.5$, $\lambda_R=0.5t$, (b) for $\alpha=1$, $\lambda_R=0.5t$, (c) for $\alpha=0.5$, $\lambda_R=0.8t$, and (d) for $\alpha=1$, $\lambda_R=0.8t$ with $R=10$ $nm$ are shown. We have taken $t=1eV$, and total angular momentum quantum numbers are denoted by $m=-1,-2$ (red curves), $m=0$ (green curves), and $m=1,2$ (blue curves).}
\label{fig:Evsphi_Rash}
\end{figure}
\end{widetext}

The energy levels as a function of the external magnetic flux ($\beta=\Phi/\Phi_0$) are depicted in Fig. \ref{fig:Evsphi_Rash} for a quantum ring with $R=10$ $nm$, considering different cases, namely, (a) $\alpha=0.5$, $\lambda_R=0.5t$, (b) $\alpha=1$, $\lambda_R=0.5t$, (c) $\alpha=0.5$, $\lambda_R=0.8t$, and (d) $\alpha=1$, $\lambda_R=0.8t$. The curves are represented by red, green, and blue colors corresponding to $m=-1$, $m=0$, and $m=1$, respectively. The magnetic field dependence of the energy spectra become evident when we rewrite Eq. (\ref{Egn_Rash_flux}) as, 
\begin{equation}
E^2_{2}-\frac{\epsilon^2}{4}\big[1+4(m+\beta)^2-4(m+\beta)\frac{1-\alpha^2}{1+\alpha^2}\big]\\
(1+\frac{\lambda_R^2}{t^2})=0
\end{equation}
and
\begin{equation}
\begin{aligned}
E^2_{3}-\frac{\epsilon^2}{4}\big[\big(1+4(m+\beta)^2\big)(1+\frac{\lambda_R^2}{t^2}\frac{1-\alpha^2}{1+\alpha^2})+\\
4(m+\beta)(\frac{\lambda_R^2}{t^2}+\frac{1-\alpha^2}{1+\alpha^2})\big]=0.
\end{aligned}
\end{equation} 
Thus, the energies display a hyperbolic dependence on the applied magnetic field, exhibiting extrema at the flux values given by $\Phi/\Phi_0=-m+\frac{1}{2}\frac{1-\alpha^2}{1+\alpha^2}$ for $\uparrow$-spin band $E_{2}$, which is independent of the strength of the Rashba coupling, but depends on the values of $m$ and the parameter $\alpha$. For the dice lattice ($\alpha=1$), the extrema occur at $\Phi/\Phi_0=-m$. However, the extrema for the $\downarrow$-spin band $E_{3}$ occur at $\Phi/\Phi_0=-m-\frac{1}{2}\frac{\frac{\lambda_R^2}{t^2}+\frac{1-\alpha ^2}{1+\alpha^2}}{1+\frac{\lambda_R^2}{t^2}\frac{1-\alpha^2}{1+\alpha^2}}$, showing a dependency on the strength of Rashba SOC, $\alpha$ and $m$. For the dice lattice, the extrema are obtained at $\Phi/\Phi_0=-m-\frac{1}{2}\frac{\lambda_R^2}{t^2}$. The energy gaps at the extrema points are given by, $\Delta E_{2}=\frac{2\epsilon\alpha}{1+\alpha^2}\sqrt{1+\frac{\lambda_R^2}{t^2}}$ and $\Delta E_{3}=\frac{2\epsilon\alpha}{1+\alpha^2}\sqrt{\frac{1-\frac{\lambda_R^4}{t^4}}{1+\frac{\lambda_R^2}{t^2}\frac{1-\alpha^2}{1+\alpha^2}}}$. Therefore, it is observed that for a fixed value of Rashba coupling, the energy gaps for both the spin bands increase with the increase in $\alpha$. However, the minimum energy gap for both the spin bands is independent of $m$. Also the $\downarrow$-spin bands, namely, $E_{3}$ have lower energy than the $\uparrow$-spin bands ($E_2$). The $\uparrow$-spin $E_2$ and $\downarrow$-spin $E_3$ bands are illustrated in the Fig. \ref{fig:Evsphi_Rash}(a). For the dice lattice case, and for $\lambda_R=0.5t$, the energy gaps are obtained as, $\Delta E_{2}\approx 74$ $meV$ and $\Delta E_{3}\approx 64$ $meV$ which can be verified from the Fig. \ref{fig:Evsphi_Rash}(b). Furthermore, from Fig. \ref{fig:Evsphi_Rash} it is evident that $E_{2}(m)\neq E_{2}(-m)$ and $E_{3}(m)\neq E_{3}(-m)$, indicating the existence of finite spin currents.

\subsubsection{Charge persistent currents}
The charge persistent current in the low-energy state can be calculated using the linear response formula, $j_Q=-\sum_{m,\kappa}\frac{\partial E}{\partial \Phi}$, where the sum refers to all (and only) the occupied states (for the valence band ($\kappa=-1$)) and the $m$ values are chosen carefully to perform the summation. Since the current is periodic in $\Phi/\Phi_0$ with a period of 1 (that is $\Phi=\Phi_0$), we restrict the discussion to the window $-1\leq \Phi/\Phi_0 \leq 1$. The analytical form for the charge current is,
\begin{widetext}
\begin{equation}
\label{per_curr_rash}
j_{Q,\lambda_R}^\kappa=-\frac{\epsilon^2\kappa}{2\Phi_0}\sum_m\frac{(1+\frac{\lambda_R^2}{t^2})\big[2(m+\frac{\Phi}{\Phi_0})-\frac{1-\alpha^2}{1+\alpha^2}\big]}{E_2(\Phi)}-\frac{\epsilon^2\kappa}{2\Phi_0}\sum_m\frac{2(m+\frac{\Phi}{\Phi_0})(1+\frac{\lambda_R^2}{t^2}\frac{1-\alpha^2}{1+\alpha^2})+(\frac{\lambda_R^2}{t^2}+\frac{1-\alpha^2}{1+\alpha^2})}{E_3(\Phi)}.
\end{equation}

\begin{figure}
\centering
\begin{minipage}{.485\linewidth}
  \includegraphics[width=\linewidth]{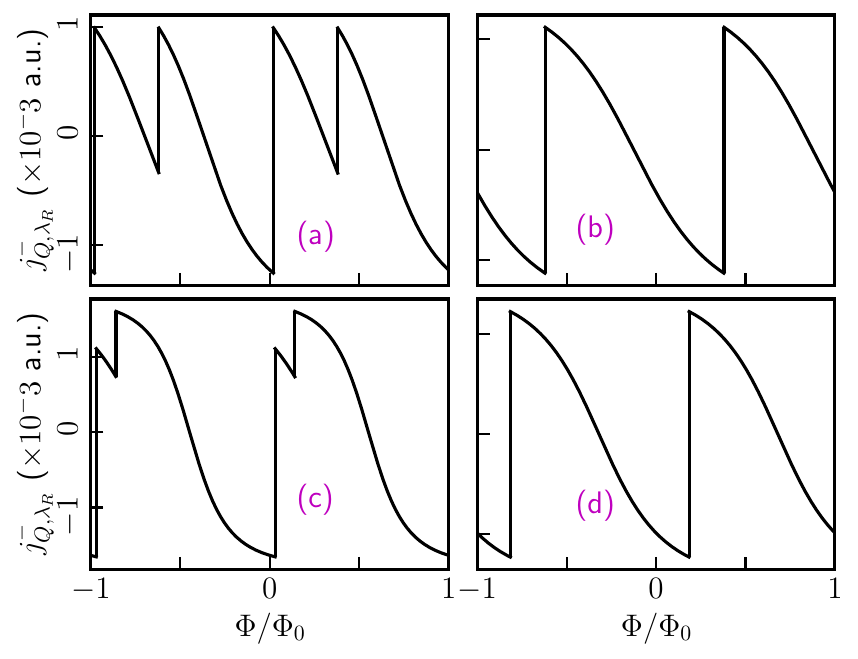}
  \caption{(Color online) The charge persistent current as a function of external magnetic flux for the low-energy state for (a) $\alpha=0.5$, $\lambda_R=0.5t$, (b) $\alpha=1$, $\lambda_R=0.5t$, (c) $\alpha=0.5$, $\lambda_R=0.8t$, and (d) $\alpha=1$, $\lambda_R=0.8t$ are shown. The ring has a radius, $R=10$ $nm$ and $t=1eV$ taken.}
  \label{fig:per_Rash}
\end{minipage}
\hspace{.01\linewidth}
\begin{minipage}{.485\linewidth}
  \includegraphics[width=\linewidth]{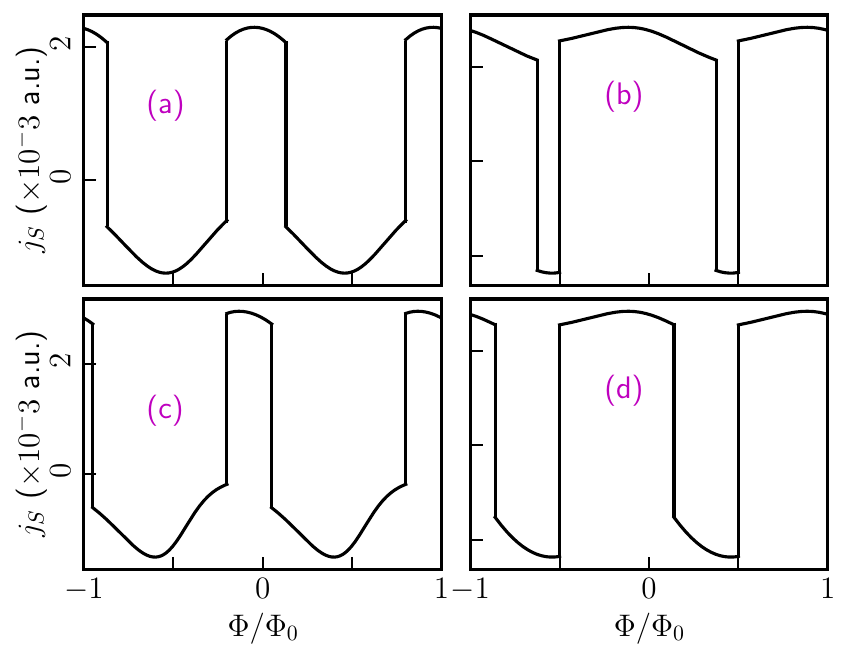}
  \caption{(Color online) The equilibrium spin current as a function of external magnetic flux for (a) $\alpha=0.5$, $\lambda_R=0.5t$, (b) $\alpha=1$, $\lambda_R=0.5t$, (c) $\alpha=0.5$, $\lambda_R=0.8t$, and (d) $\alpha=1$, $\lambda_R=0.8t$ are shown. The ring has a radius of $R=10$ $nm$ and $t=1eV$ taken.}
  \label{fig:spin_cur_Rash}
\end{minipage}
\end{figure}
\end{widetext}

The spin branches closest to the Fermi energy exhibit non-monotonic behaviour, resulting in two distinct contributions to the charge current coming from the $\uparrow$-spin and $\downarrow$-spin components. Since we are calculating the current contributions arising from the low-energy states, it is clear from Fig. \ref{fig:Evsphi_Rash} that for a certain range of $\Phi/\Phi_0$, only one energy state labelled by a particular value of $m$ is present. Hence, the sum in Eq. (\ref{per_curr_rash}) comprises of only one value of $m$. We computed the current taking the contributions from both the spin branches and the results are depicted in Fig. \ref{fig:per_Rash} for (a) $\alpha=0.5$, $\lambda_R=0.5t$, (b) $\alpha=1$, $\lambda_R=0.5t$, (c) $\alpha=0.5$, $\lambda_R=0.8t$, and (d) $\alpha=1$, $\lambda_R=0.8t$, for a particular ring radius, $R=10nm$. The asymmetric structure of the spectral features between the two spin branches allows for the possibility of a net spin currents, as we shall see below. For all values of $\alpha$, the persistent current oscillates periodically with $\Phi/\Phi_0$, with a periodicity of $\Phi/\Phi_0=1$. Fig. \ref{fig:per_Rash} illustrates that the persistent current can be tuned by adjusting the parameter $\alpha$ for a fixed value of the Rashba coupling ($\lambda_R$). Moreover, the charge persistent current can be manipulated via $\lambda_R$ for a fixed $\alpha$, since the Rashba parameter can be controlled by a gate voltage. In contrast, the intrinsic spin-orbit coupling ($\tilde{\lambda}$) cannot be tuned by applying external fields. However, experimental evidence has shown that the intrinsic spin-orbit coupling strength can still be manipulated via carbon hybridization technique and may contribute in tuning the charge persistent current[\onlinecite{Bala}].

\subsubsection{Equilibrium spin currents}
We shall now study equilibrium spin currents. In contrast to the formalism for obtaining the charge current, one can obtain the spin currents by accounting for distinct velocities for different spin branches. Thus, we define equilibrium spin current as,
\begin{equation}
j_S=j_Q(\uparrow)-j_Q(\downarrow).
\label{Eq.spin_cur}
\end{equation}
We have calculated the equilibrium spin currents following the procedure discussed earlier. The peculiar separation of the spin branches results in velocity differences between the two spin projections, giving rise to a spin current, as shown in Fig. \ref{fig:spin_cur_Rash}. The figure illustrates a significant spin current for small values of the flux, which can be attributed to the large charge current originating from a single spin branch.

The striking feature is that the magnitude as well as the pattern of the spin currents depend upon the parameters $\alpha$ and the strength of the Rashba coupling ($\lambda_R$). We present results for (a) $\alpha=0.5$, $\lambda_R=0.5t$, (b) $\alpha=1$, $\lambda_R=0.5t$, (c) $\alpha=0.5$, $\lambda_R=0.8t$, and (d) $\alpha=1$, $\lambda_R=0.8t$ for a ring of radius, $R=10$ $nm$. The presence of the Rashba coupling breaks inversion symmetry (in addition to the $\sigma_z$ symmetry) in the plane even for small $\lambda_R$. The symmetry breaking determines the spin labelling of the energy branches that take part in yielding the spin currents. Additionally, the spin currents exhibit periodic behaviour with $\Phi/\Phi_0$, with a periodicity equal to one flux quantum.

\section{Pseudospin-1 QR with ISOC in addition with RSOC}
\label{Kane}
We add the Rashba SOC term to the Hamiltonian in Eq. (\ref{soc}) of the pseudospin-1 fermionics. Thus, we are dealing with a Kane-Mele Hamiltonian with a Rashba term for pseudospin-1 $\alpha$-$T_3$ QR. The resulting Hamiltonian is,
\begin{widetext}
\begin{equation}
\label{Ham_Kane}
\begin{split}
H = t\sum_{\langle ij \rangle} c_{i\sigma}^\dagger c_{j\sigma}+\alpha t\sum_{\langle ik\rangle} c_{i\sigma}^\dagger c_{k\sigma}+\frac{i\lambda}{3\sqrt{3}}\sum_{\langle \langle ij\rangle\rangle\sigma\sigma^\prime}\nu_{ij}c^\dagger_{i\sigma}\sigma_zc_{j\sigma^\prime}+\frac{i\alpha\lambda}{3\sqrt{3}}\sum_{\langle \langle ik\rangle\rangle\sigma\sigma^\prime}\nu_{ik}c^\dagger_{i\sigma}\sigma_zc_{k\sigma^\prime}+\lambda_R\sum_{\langle ij\rangle\sigma\sigma^\prime}ic_{i\sigma}^\dagger(\hat{D}_{ij}\cdot\vec{\tau})_{\sigma\sigma^\prime}c_{j\sigma^\prime}\\
+\alpha\lambda_R\sum_{\langle ik\rangle\sigma\sigma^\prime}ic_{i\sigma}^\dagger(\hat{D}_{ik}\cdot\vec{\tau})_{\sigma\sigma^\prime}c_{k\sigma^\prime}+\bigg(h.c.\bigg)
\end{split}
\end{equation}
\end{widetext}
All the terms and notations have described earlier in Sec. \ref{Hal} and Sec. \ref{Rash}.
\subsection{Results and discussions}
\subsubsection{No magnetic field}
\begin{widetext}

\begin{figure}[h!]
\centering
\includegraphics[width=16cm, height=10cm]{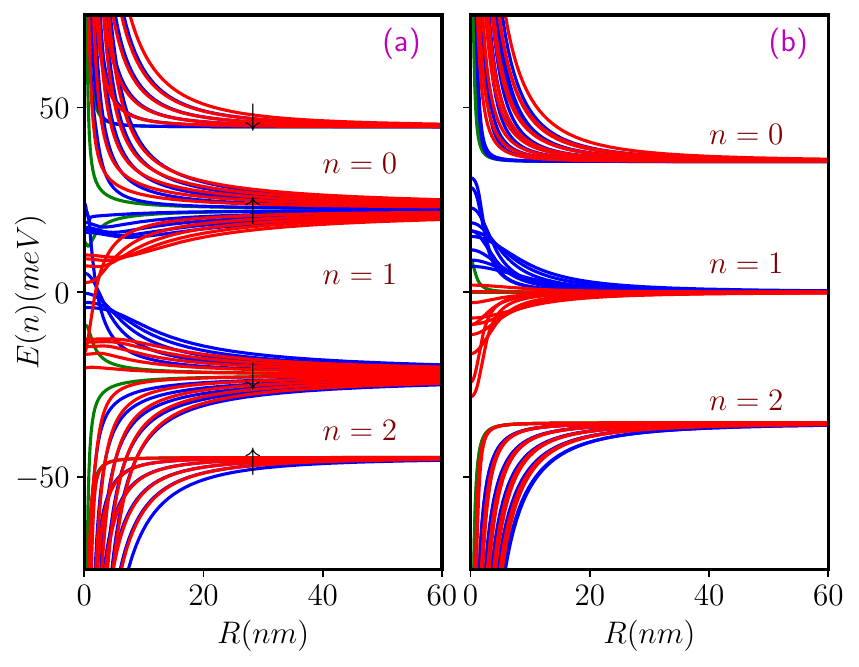}
\caption{(Color online) The energy spectra as a function of the ring radius $R$ of the $\alpha$-$T_3$ quantum ring at ${\bf K}$ valley for (a) $\alpha=0.5$, and  (b) $\alpha=1$. $n=0,1,2$ denote the CB, distorted FB, and VB respectively. The parameters are taken as $\tilde{\lambda}=0.05t$, $\lambda_R=0.5t$ and $t=1eV$.}
\label{fig:Kane_EvsR}
\end{figure}
\end{widetext}
The electronic energy spectra at the ${\bf K}$-valley as a function of the ring radius $R$ of the system, via varying both $\lambda$ and $\lambda_R$, have been obtained via numerical diagonalization of Eq. (\ref{Ham_Kane}) and are shown in Fig. \ref{fig:Kane_EvsR}. When both the Rashba coupling ($\lambda_R$) and intrinsic spin-orbit coupling ($\tilde{\lambda}$) are considered, the energy bands for spin-$\uparrow$ and spin-$\downarrow$ states become distinguishable for any values of $m$. Consequently, in this case we get a pair of conduction ($n=0$), distorted flat ($n=1$), and valence ($n=2$) bands for each spin. In Fig. \ref{fig:Kane_EvsR}, we have plotted the energy spectra for two different $\alpha$ values, namely, $\alpha=0.5$ (Fig. \ref{fig:Kane_EvsR}(a)) and $\alpha=1$ (Fig. \ref{fig:Kane_EvsR}(b)) with $m=0$ (green curves), $1$ (blue curves), and $-1$ (red curves). The parameters used in the calculations are $\tilde{\lambda}=0.05t$ and $\lambda_R=0.5t$. Notably, the energy levels are now non-degenerate for all values of $\alpha$. It is worth mentioning that for all values of $\alpha$, and for small radii, all the branches of the VB and CB vary as $\sim 1/R$. Within the CB and VB, the energy splitting between the spin bands corresponding to a specific $m$ value increases with $\alpha$ in the range namely, $0<\alpha\leq 0.5$. However, as $\alpha$ increases within the range $0.5<\alpha\leq 1$, the energy separation between the spin bands decreases. Additionally, regardless of the value of $\alpha$, the energy separation diminishes as the magnitude of $m$ ($|m|$) increases. Within the distorted $n=1$ band, the degree of distortion increases as $\alpha$ increases in the small radius limit. A finite gap exists between the $\uparrow$- and $\downarrow$-spin bands of the $m=0$ band. This gap increases with $\alpha$ in the range $0<\alpha\leq 0.5$, while it decreases in the range $0.5<\alpha\leq 1$. This non-monotonicity is an interesting feature of our study. Furthermore, the energy separation between the $\uparrow$- and $\downarrow$-spin bands decreases as $\alpha$ and $|m|$ increases. Moreover, certain spin bands intersect at a particular value of the radius of the ring, as depicted in Fig. \ref{fig:Kane_EvsR}(a). As we increase $\alpha$, the crossing point gradually shifts towards larger values of $R$. In the limit $\alpha=1$, the $\downarrow$-spin band of $m=0$ remains flat at zero energy, and the spin bands converge at large values of $R$ as shown in Fig. \ref{fig:Kane_EvsR}(b). Furthermore, in the regime of large radii, we observe intriguing features in the energy spectra, which we shall discuss below.

As we increase $\alpha$ in the range $0<\alpha\leq 0.5$, a distinct separation between the spin branches of a specific band (VB, distorted FB, or CB) becomes evident, and this separation becomes more prominent as $\alpha$ increases. Specifically, the $\uparrow$-spin bands of the distorted flat band ($n=1$) migrate away from the band center ($E=0$) and move closer to the $\uparrow$-spin bands of the conduction band ($n=0$).  The $\uparrow$-spin energy bands of $n=1$ band (distorted FB) merge to a value of $\alpha\tilde{\lambda}$, while the $\uparrow$-spin bands of the $n=0$ band converge to a value given by $(1-\alpha)\tilde{\lambda}$ in the limit of large $R$. It is important to note that the specific criteria for considering $R$ as "large" vary, in a sense they depend upon the value of $m$. Conversely, the $\downarrow$-spin bands of $n=1$ shift towards the $\downarrow$-spin bands of $n=2$ (VB). Again they merge to a value $-\alpha\tilde{\lambda}$ and $-(1-\alpha)\tilde{\lambda}$ respectively (for more detailed information, please refer to the Fig. \ref{fig:appc} in Appendix C). Fig. \ref{fig:Kane_EvsR}(a) illustrates the results for $\alpha=0.5$ where $\uparrow$-spin bands of $n=0$ and $n=1$ merge to a value $\tilde{\lambda}/2$, while the $\downarrow$-spin bands merge to $-\tilde{\lambda}/2$. Additionally, the $\downarrow$-spin bands corresponding to $n=0$ merge to $\tilde{\lambda}\cos\xi$ in the large $R$ limit, while the $\uparrow$-spin bands of the $n=2$ merge to $-\tilde{\lambda}\cos\xi$ as depicted  in Fig. \ref{fig:Kane_EvsR}(a). For $\alpha=0.5$, $\uparrow$-spin branches of the VB ($n=2$) merge to $-45meV$, while the $\downarrow$-spin branches of the CB ($n=0$) merge to $45meV$ which match very well with our analytic results discussed above.

For the range $0.5<\alpha\leq 1$, as we increase $\alpha$ the separation between the spin branches for a specific band (VB, distorted FB or CB) diminishes, eventually leading to spin-mixed bands at $\alpha=1$. In contrast to the previous scenario, in the current case, here the $\uparrow$-spin bands of the distorted flat band ($n=1$) merge together at a value $(1-\alpha)\tilde{\lambda}$, while the $\downarrow$-spin bands merge to $-(1-\alpha)\tilde{\lambda}$ in the limit of large $R$. On the other hand, the $\uparrow$-spin and the $\downarrow$-spin branches of the $n=0$ band merge to $\alpha\tilde{\lambda}\cos\xi$ and $\tilde{\lambda}\cos\xi$ respectively, while the $\uparrow$-spin and the $\downarrow$-spin branches of the $n=2$ band merge to $-\tilde{\lambda}\cos\xi$ and $-\alpha\tilde{\lambda}\cos\xi$ respectively. Fig. \ref{fig:Kane_EvsR}(b) displays the results for the $\alpha=1$ case where the spin bands of the $n=1$ band merge together at zero energy and the spin bands of the $n=0$ and $n=2$ merge to $\pm \tilde{\lambda}/\sqrt{2}$ respectively, which are in accordance with the aforementioned explanations.

\begin{widetext}

\begin{figure}[h!]
\centering
\includegraphics[width=16cm, height=9cm]{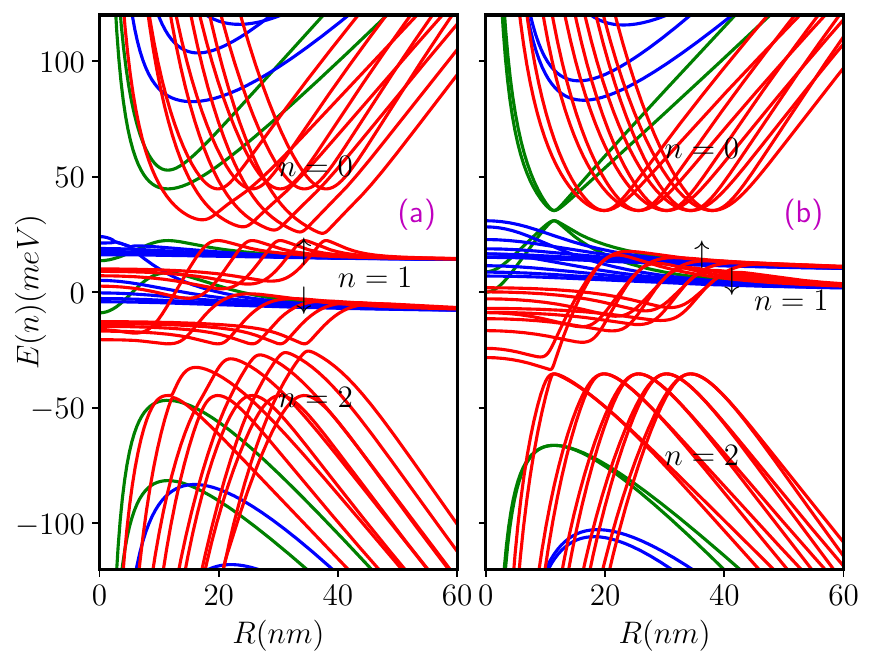}
\caption{(Color online) The energy spectra as a function of ring radius $R$ of the $\alpha$-$T_3$ quantum ring in presence of an external magnetic field of $B-0=5T$ at ${\bf K}$-valley for (a) $\alpha=0.5$, and  (b) $\alpha=1$. $n=0,1,$ and $2$ denote the CB, distorted FB, and VB respectively. The parameters are taken as $\tilde{\lambda}=0.05t$, $\lambda_R=0.5t$ and $t=1eV$.}
\label{fig:Kane_EvR_B}
\end{figure}
\end{widetext}

\subsubsection{In presence of magnetic field}
Now, let us consider the case when the ring is subjected to a perpendicular magnetic field. The field significantly alters the above scenario. In Fig. \ref{fig:Kane_EvR_B}, we show the dependence of several energy levels on the ring radius ($R$) for two cases namely, $\alpha=0.5$ and $\alpha=1$, assuming a magnetic field strength of $B_0=5$T. Each level exhibits a non-monotonic behaviour as a function of $R$. Notably, the energy levels of the CB ($n=0$) and VB ($n=2$) attain inflection points (minimum for CB and maximum for VB) at specific values of $R$. However, the locations of these points depend on the value of $m$,  as shown in Fig. \ref{fig:Kane_EvR_B}. In the limit of small $R$, all the energy levels vary inversely with $R$. On the other hand, the energy scales approximately linearly with $R$, in the limit of large $R$. It is worth mentioning that the criteria for considering $R$ as "large" differ for different values of $m$. Furthermore, within the CB ($n=0$) and VB ($n=2$) for a particular value of $m$ and $\alpha$, the energy splitting between the spin branches decreases as $R$ increases before reaching the inflection points. However, after reaching these points, the energy splitting between the spin bands increases with $R$ (see Fig. \ref{fig:Kane_EvR_B}(a)). Additionally, the energy splitting between the bands decreases as $|m|$ increases, for all values of $\alpha$. Moreover, for a specific $|m|$ value, the energy splitting decreases as $\alpha$ increases, and eventually, the spin bands merge at the extremum point in the limit of $\alpha=1$ as shown in the Fig. \ref{fig:Kane_EvR_B}(b).

Let us briefly discuss the behaviour of energy levels within the distorted FB ($n=1$) considering to the choice of parameters $m$, $R$, and $\alpha$. In the small $R$ limit, and for a specific value of $m$, increasing $\alpha$ results in a greater degree of distortion within the distorted FB. Additionally, for a fixed value of $\alpha$, the distortion also increases with $|m|$ increase. Now, focusing on a fixed value of $|m|$, within the range of $0 < \alpha \leq 0.5$, an increase in $\alpha$ leads to an increase in the separation between the spin-$\uparrow$ and spin-$\downarrow$ bands. However, within the range of $0.5 < \alpha \leq 1$, the splitting between the spin bands decreases as $\alpha$ increases. Moreover, for a particular $\alpha$ value, in the small $R$ limit, the energy separation increases with increasing $|m|$. On the other hand, in the large R limit, the spin-$\uparrow$ and spin-$\downarrow$ bands with different $m$ values merge with each other, resulting in spin-split energy bands, which happens for any value of $\alpha$. However, in the large $R$ limit, the splitting between the spin bands increases as $\alpha$ increases in the other range, that is, $0< \alpha \leq 0.5$, while the separation between the bands decreases as $\alpha$ increases within the range of $0.5 < \alpha \leq 1$. For more detailed information, please refer to the Fig. \ref{fig:appd} of Appendix D.

In Fig. \ref{fig:Kane_EvR_B}(a), we illustrate the case for $\alpha=0.5$. Here, for a fixed $|m|$, the energy splitting between the spin bands increases as $R$ increases. The energy separation between the spin bands increases as $|m|$ increases in the small $R$ limit. The $\uparrow$-spin bands of $n=1$ shift towards the $\uparrow$-spin bands of $n=0$, and as $m$ increases in the negative direction, they also move closer to each other. Similarly, the $\downarrow$-spin bands shift towards the $\downarrow$-spin bands of the $n=2$, and as negative values of $m$ increases, they move closer to each other. In Fig. \ref{fig:Kane_EvR_B}(b), we depict the case for $\alpha=1$. Here, the separation between the spin bands of $m=0$ and negative values of $m$ decreases with increasing $R$. Further, at a certain radius (inflection points), they touch each other. Subsequently, the band gap increases, leading to the emergence of spin-separated bands in the large $R$ limit. However, for positive values of $m$ no band touching occurs. Furthermore, the bands of the distorted FB with $m = 0$ shift towards the bands of CB with $m=0$, and as the negative values of $m$ increases, the shifting points move further apart. Conversely, the spin-$\downarrow$ band of the distorted FB with $m = -1$ shifts towards the bands of VB with $m = -1$, and as negative values of $m$ increases, the shifting points also move further apart.

\subsubsection{Charge persistent currents}
\begin{figure}[h!]
\centering
\includegraphics[width=\linewidth, height=0.8\linewidth]{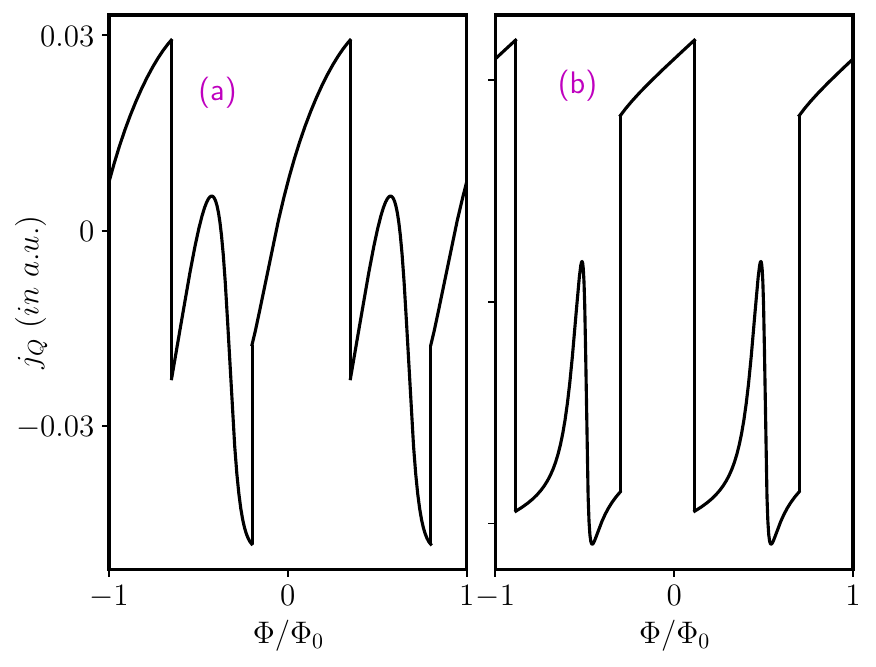}
\caption{(Color online) The charge persistent current as a function of external magnetic flux for the low-energy states in presence of both the ISOC and RSOC terms for (a) $\alpha=0.5$ and (b) $\alpha=1$. The radius of the ring is $10$ $nm$ and the parameters are taken as $\tilde{\lambda}=0.05t$, $\lambda_R=0.5t$ and $t=1eV$.}
\label{fig:Kane_per_curr}
\end{figure}
From Fig. \ref{fig:Kane_EvR_B}, it is evident that the presence of an external field in this case causes the flat band to no longer remain flat, which in turn contributes to the transport properties of the system. To investigate the impact of the distorted flat band, we calculate the persistent current. The charge persistent current in the low-energy state can be determined using the linear response definition, $j_Q=-\sum_{m}\frac{\partial E}{\partial \Phi}$, where the summation is over the low-energy occupied states only. We confined our discussion in the range $-1\leq \Phi/\Phi_0 \leq 1$, where the occupied states exist in the valence band and the distorted flat band. We follow the same procedure as stated in Secs. \ref{Hal} and \ref{Rash}. The persistent current then obtained as a combination of contributions from the valence band ($n=2$) and the distorted flat band ($n=1$) as, $j_Q=j_Q^{n=1}+j_Q^{n=2}$. It is noteworthy that the distortion of the energy levels in the flat band gives rise to finite persistent current, in contrast to the case of Rashba SOC (in Sec. \ref{Rash}). In Fig. \ref{fig:Kane_per_curr}, we demonstrate the variation of the persistent current with $\Phi/\Phi_0$, for $\tilde{\lambda}=0.05t$, $\lambda_R=0.5t$, and $R=10nm$ corresponding to two different cases, namely $\alpha=0.5$ (Fig. \ref{fig:Kane_per_curr}(a)), and $\alpha=1$ (Fig. \ref{fig:Kane_per_curr}(b)). The introduction of $\tilde{\lambda}$ and $\lambda_R$ completely changes the oscillation pattern of the persistent current from the case of $\tilde{\lambda} \neq 0$, $\lambda_R=0$ (see Fig. \ref{fig:valley_cur}) and $\tilde{\lambda}=0$, $\lambda_R \neq 0$ (see Fig. \ref{fig:per_Rash}). Furthermore, the current exhibits periodic oscillations in $\Phi/\Phi_0$ with the periodicity $\Phi/\Phi_0 =1$. The period of oscillation remains insensitive to the value of $\alpha$. However, the oscillation pattern varies for different $\alpha$ values. Moreover, the charge persistent current can be manipulated by $\lambda_R$, which is tunable via an external.

\subsubsection{Equilibrium spin currents}
\begin{figure}[h!]
\centering
\includegraphics[width=\linewidth, height=0.8\linewidth]{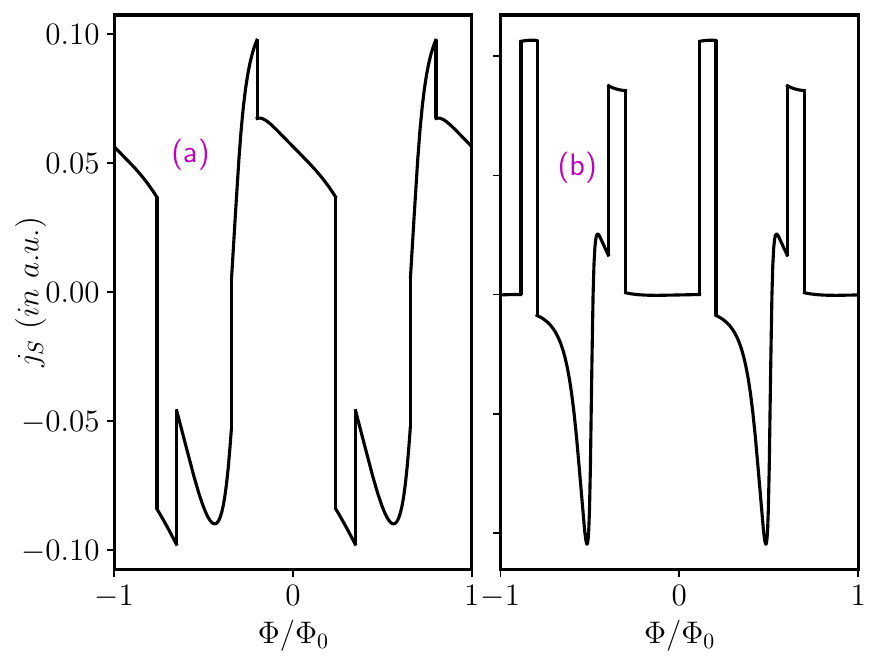}
\caption{(Color online) The equilibrium spin current as a function of external magnetic flux for the low-energy states considering both the ISOC and RSOC terms for (a) $\alpha=0.5$ and (b) $\alpha=1$. Again the ring radius is $10$ $nm$ and the parameters are taken as $\tilde{\lambda}=0.05t$, $\lambda_R=0.5t$ and $t=1eV$.}
\label{fig:Kane_per_curr_spin}
\end{figure}
Finally, we  consider the equilibrium spin currents. We define the equilibrium spin current as in Eq. (\ref{Eq.spin_cur}), which combines charge current contributions from different spin-split labels. We have calculated this taking into account of the low-energy states of the separate spin branches. In this case, the inversion symmetry is broken inside the plane, leading to asymmetric spin branches. The peculiar separation of the spin branches results in velocity differences between the two spins, leading to a spin current, as illustrated in Fig. \ref{fig:Kane_per_curr_spin}. We have consider two cases, $\alpha=0.5$ (Fig. \ref{fig:Kane_per_curr_spin}(a)) and $\alpha=1$ (Fig. \ref{fig:Kane_per_curr_spin}(b)), corresponding to the parameter values $\tilde{\lambda}=0.05t$, $\lambda_R=0.5t$, and $R=10nm$. The figures depict a large spin current for small fluxes, which can be attributed to the large charge currents originating from a single spin branch. The current oscillates periodically in $\Phi/\Phi_0$, with the periodicity $\Phi=\Phi_0$. Furthermore, the pattern of oscillation varies for different $\alpha$ values, while the oscillation period remains independent of the values of $\alpha$. The Rashba coupling breaks the inversion symmetry in the plane, which determines the spin labelling of the energy branches participating in the spin current. Additionally, the intrinsic SOC term makes the flat band dispersive, allowing it to contribute to the equilibrium spin currents. Furthermore, the equilibrium spin current can also be manipulated by Rashba coupling and ISOC, as they are tunable via external field and edge functionalization.

\section{Summary and conclusions}
\label{Sum}
We have thoroughly examined the electronic properties of the $\alpha$-$T_3$ pseudospin-1 fermionic quantum ring in presence of an external magnetic field, including the effects of Rashba and intrinsic spin-orbit couplings separately, as well as their combined effects. Our Hamiltonian for a circular ring possesses four quantum numbers that are employed to describe the energy eigenvalues, namely, the valley index $\zeta$, the band index $n$, the spin quantum number $\sigma$ for labelling the spin, and the angular momentum quantum number $m$. Confinement of the carriers in the ring leads to quantization of the energy levels. The key observations from our analyses include the following.

When only intrinsic spin-orbit coupling is present ($\tilde{\lambda}\neq 0$, $\lambda_R=0$) and in the absence of a magnetic field, the flat band becomes dispersive for small ring radii, except for the $m = 0$ band corresponding to $\alpha = 1$ (dice limit). At large radii, the energy levels in the valence band (VB) and the conduction band (CB) converge to specific energy values determined by the parameters $\tilde{\lambda}$ and $\alpha$. The introduction of the magnetic field makes all the levels in the flat band dispersive for all finite values of $\alpha$ ($\alpha\neq 0$). The energy levels in the VB and CB display significant deviations from their typical $R$ dependence, showing $\sim 1/R$ behaviour for small $R$ and linear $R$ behaviour for large $R$. The persistent currents exhibit $\Phi_0$ periodic oscillations at the individual valleys, resembling Aharonov-Bohm oscillations. The valley currents, derived by combining charge current contributions from the two valleys, exist only for intermediate to graphene (pseudospin-1/2) and dice lattice (pseudospin-1), $0 < \alpha < 1$, while in the limiting cases i.e., $\alpha=0$ and $\alpha=1$, they vanish. Equilibrium spin currents obtained from combining the charge current contributions from different spin labels also exist for all values of  $\alpha\neq 0$ (pseudospin-1 system) and exhibit $\Phi_0$ periodic oscillations, with the oscillation pattern depending on $\alpha$ and $\tilde{\lambda}$.

As a second scenario, we have considered only Rashba spin-orbit coupling present ($\tilde{\lambda}=0$, $\lambda_R\neq 0$). The inclusion of $\lambda_R$ leads to six bands in the spectrum, including two non-dispersive flat bands and four dispersive spin-split VB and CB. The flat band consists of a large number of degenerate levels at zero-energy, which are insensitive to the applied magnetic field. In the absence of a magnetic field, all the energy levels in the CB and VB exhibit inverse dependence on the ring radius $R$, and are independent of $\alpha$. Interestingly, the $\uparrow$-spin energy levels corresponding to $\alpha=1$ is two-fold degenerate, except for the $m=0$ level. However, the $\downarrow$-spin bands are non-degenerate for all values of $\alpha$. When the ring is subjected to a perpendicular magnetic field, the energy levels deviate significantly from their typical $R$-dependence, displaying $\sim 1/R$ behaviour for small $R$ and $\sim R$ behaviour for large $R$. The persistent currents show $\Phi_0$ periodic oscillations, with distinct patterns for different $\alpha$ and $\lambda_R$ values. We derived equilibrium spin currents, by combining the charge current contributions from different spin branches. Equilibrium spin currents are also present for all the values of $\alpha$ and display the same periodic behaviour.

As a final case, we have considered the effects of both spin-orbit couplings, namely, $\tilde{\lambda}\neq 0$ and $\lambda_R\neq 0$. In the absence of a magnetic field, all the flat bands become dispersive for small ring radii, except for the $m = 0$ $\downarrow$-spin band corresponding to $\alpha = 1$. The introduction of a magnetic field leads to all the levels of the flat band becoming dispersive for all $\alpha\neq 0$, with the VB and CB behaving with the ring radius $R$ similar to the previous cases. Further, the persistent currents exhibit $\Phi_0$ periodic oscillations with a pattern distinct from the previous cases. Equilibrium spin currents are non-zero as well for all values of $\alpha$ and display similar periodic behaviour with a distinct oscillation pattern. The parameters $\alpha$, $\tilde{\lambda}$, and $\lambda_R$ are tunable in our work ($\tilde{\lambda}$, $\lambda_R$ are experimentally tunable too), the persistent currents can be controlled by varying these parameters.

In summary, we have investigated the properties of the $\alpha$-$T_3$ pseudospin-1 fermionic quantum ring in details. Our studies include energy spectrum, persistent currents, and size dependencies, and they depend upon the spin-orbit couplings and the magnetic field. By tuning the parameters $\alpha$, $\tilde{\lambda}$, and $\lambda_R$, we can manipulate the persistent currents, making it a controllable feature in our system.

\section*{ACKNOWLEDGMENTS}
One of the authors MI sincerely acknowledge Dr. Tutul Biswas for fruitful
discussions.

\begin{widetext}
\section*{Appendix A: Evolution of the energy spectra for intrinsic SOC}
\label{AppA}

\begin{figure}[H]
\centering
\includegraphics[width=12cm, height=8.9cm]{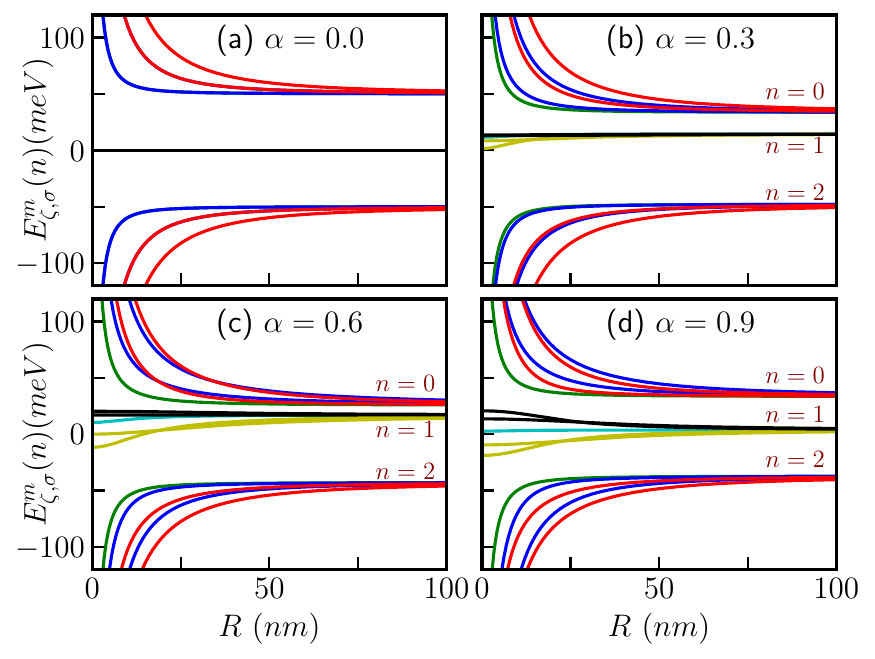}
\caption{(Color online) Evolution of the spin-$\uparrow$ and ${\bf K}$-valley energy dispersion as a function of $R$ of the $\alpha$-$T_3$ QR for different values of $\alpha$. The parameters are taken as $\tilde{\lambda}=0.05t$ and $t=1eV$. Panels (a), (b), (c), and (d) denote $\alpha=0,0.3,0.6$, and $0.9$ respectively.}
\label{fig:appa}
\end{figure}
\section*{Appendix B: Energy spectra in presence of magnetic field for intrinsic SOC}
\begin{figure}[H]
\centering
\includegraphics[width=11.5cm, height=8.7cm]{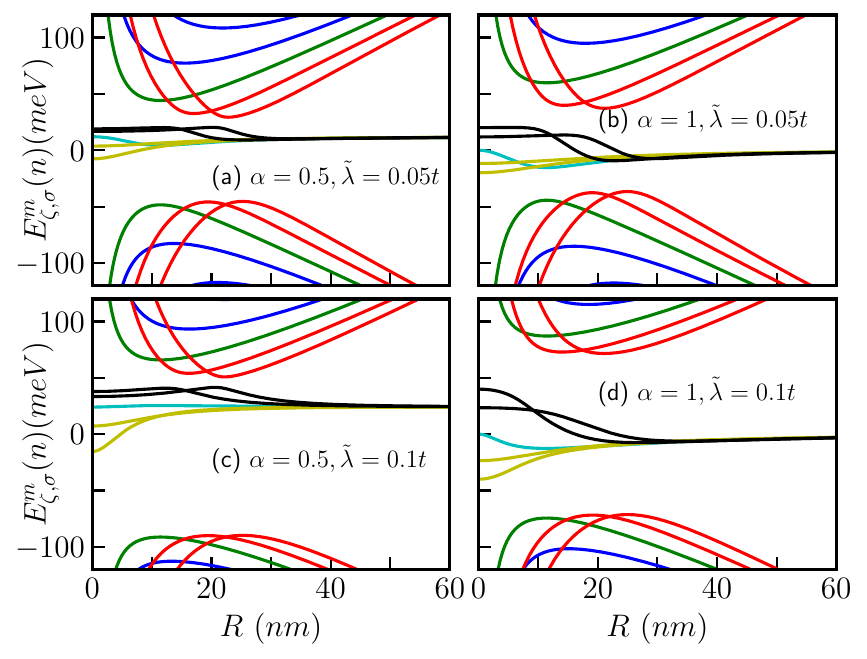}
\caption{(Color online) The energy spectra as a function of ring radius $R$ of the $\alpha$-$T_3$ quantum ring at ${\bf K}$-valley in presence of external magnetic field of $B_0=5T$ for (a) $\alpha=0.5$, $\tilde{\lambda}=0.05t$, (b) $\alpha=1$, $\tilde{\lambda}=0.05t$, (c) $\alpha=0.5$, $\tilde{\lambda}=0.1t$, and (d) $\alpha=1$, $\tilde{\lambda}=0.1t$. $n=0,1,2$ denote the CB, distorted FB, and VB respectively.}
\label{fig:appb}
\end{figure}
\section*{Appendix C: Evolution of the energy spectra for pseudospin-1 $\alpha$-$T_3$ QR with both ISOC and RSOC}
\begin{figure}[H]
\centering
\includegraphics[width=12cm, height=8.7cm]{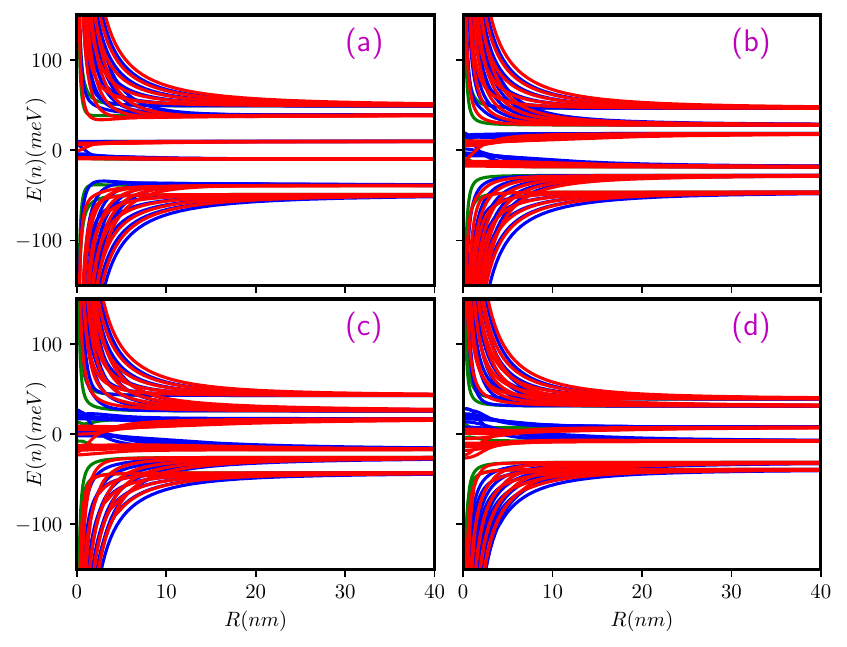}
\caption{(Color online) Evolution of the energy spectra as a function of ring radius $R$ of the $\alpha$-$T_3$ quantum ring at ${\bf K}$-valley in presence of both the ISOC and RSOC terms together for (a) $\alpha=0.2$, (b) $\alpha=0.4$, (c) $\alpha=0.6$, and (d) $\alpha=0.8$. The parameters $\tilde{\lambda}=0.05t$ and $\lambda_R=0.5t$. $n=0,1,2$ denote the CB, distorted FB, and VB respectively.}
\label{fig:appc}
\end{figure}

\section*{Appendix D: Energy spectra in presence of magnetic field for pseudospin-1 $\alpha$-$T_3$ QR with both ISOC and RSOC}

\begin{figure}[H]
\centering
\includegraphics[width=12cm, height=8.7cm]{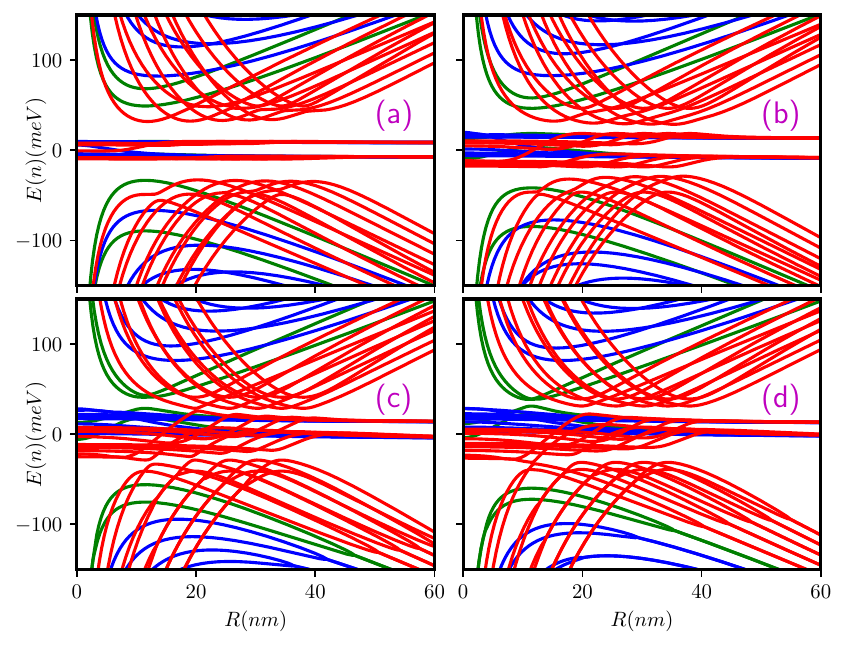}
\caption{(Color online) Evolution of the energy spectra as a function of ring radius $R$ of the $\alpha$-$T_3$ quantum ring at ${\bf K}$-valley in presence of external magnetic field of $B_0=5$T considering both the ISOC and RSOC terms together for (a) $\alpha=0.2$, (b) $\alpha=0.4$, (c) $\alpha=0.6$, and (d) $\alpha=0.8$. The parameters $\tilde{\lambda}=0.05t$ and $\lambda_R=0.5t$. $n=0,1,2$ denote the CB, distorted FB, and VB respectively.}
\label{fig:appd}
\end{figure}
\end{widetext}


\begin{thebibliography}{55}
\bibitem{But}
    M. Buttiker, Y. Imry, and R. Landauer, \href{https://doi.org/10.1016/0375-9601(83)90011-7}{Phys. Lett. A {\bf96}, 365 (1983)}.
\bibitem{Che}
    H. F. Cheung, Y. Gefen, E. K. Riedel, and W. H. Shih, \href{https://doi.org/10.1103/PhysRevB.37.6050}{Phys. Rev. B {\bf37}, 6050 (1988)}.
    \bibitem{Che1}
     H. F. Cheung, Y. Gefen, and E. K. Riedel, \href{https://doi.org/10.1147/rd.323.0359}{IBM J. Res. Dev. {\bf32}, 359 (1988)}.
     \bibitem{Che2}
     H. F. Cheung, E. K. Riedel, and Y. Gefen, \href{https://doi.org/10.1103/PhysRevLett.62.587}{Phys. Rev. Lett. {\bf62}, 587 (1989)}.
    \bibitem{Le}
    L. P. Levy, G. Dolan, J. Dunsmuir, and H. Bouchiat, \href{https://doi.org/10.1103/PhysRevLett.64.2074}{Phys. Rev. Lett. {\bf64}, 2074 (1990)}.
    \bibitem{Mon}
    G. Montambaux, H. Bouchiat, D. Sigeti, and R. Friesner, \href{https://doi.org/10.1103/PhysRevB.42.7647}{Phys. Rev. B {\bf42}, 7647 (R) (1990)}.
    \bibitem{Cha}
    V. Chandrasekhar, R. A. Webb, M. J. Brady, M. B. Ketchen, W. J. Gallagher, and A. Kleinsasser, \href{https://doi.org/10.1103/PhysRevLett.67.3578}{Phys. Rev. Lett. {\bf67}, 3578 (1991)}.
    \bibitem{Avi}
   Y. Avishai, Y. Hatsugai, and M. Kohmoto, \href{https://doi.org/10.1103/PhysRevB.47.9501}{Phys. Rev. B {\bf47} 9501 (1993)}.
   \bibitem{Bou}
   G. Bouzerar, D. Poilblanc, and G. Montambaux, \href{https://doi.org/10.1103/PhysRevB.49.8258}{Phys. Rev. B {\bf49}, 8258 (1994)}.
   \bibitem{Mai}
    D. Mailly, C. Chapelier, and A. Benoit, \href{https://doi.org/10.1103/PhysRevLett.70.2020}{Phys. Rev. Lett. {\bf70}, 2020 (1993)}.
   \bibitem{Lor}
A. Lorke, R. J. Luyken, A. O. Govorov, J. P. Kotthaus, J. M. Garcia, and P. M. Petroff, \href{https://doi.org/10.1103/PhysRevLett.84.2223} {Phys. Rev. Lett. {\bf84}, 2223 (2000)}.
   \bibitem{Alf}
   A. B. Alfonso and A. Latge, \href{https://doi.org/10.1103/PhysRevB.61.15887}{Phys. Rev. B {\bf 61}, 15887 (2000)}.
   
\bibitem{Fu}
   A. Fuhrer, S. Luscher, T. Ihn, T. Heinzel, K. Ensslin, W. Wegscheider, and M. Bichier, \href{https://doi.org/10.1038/35101552}{Nature (London) {\bf413}, 822 (2001)}.
   \bibitem{Cli}
   J. I. Climente, J. Planelles, and J. L. Movilla, \href{https://doi.org/10.1103/PhysRevB.70.081301}{Phys. Rev. B {\bf70}, 081301 (R) (2004)}.
   \bibitem{Per}
   Y. V. Pershin and C. Piermarocchi, \href{https://doi.org/10.1103/PhysRevB.72.245331}{Phys. Rev. B {\bf72}, 245331 (2005)}.
   \bibitem{Qi}
   X. Li, L. Qi, W. Guo, J. Yu, Y. Zhao, D. Cai, S. Yin, and L. H. Mao, \href{https://doi.org/10.1007/s11434-009-0294-0}{Chin. Sci. Bull. {\bf54}, 3716 (2009)}.
   \bibitem{Chav}
   A. Chaves, G. A. Farias, F. M. Peeters, and B. Szafran, \href{https://doi.org/10.1103/PhysRevB.80.125331}{Phys. Rev. B {\bf80}, 125331 (2009)}.
   \bibitem{Na}
   F. Nagasawa, D. Frustaglia, H. Saarikoski, K. Richter, and J. Nitta, \href{https://doi.org/10.1038/ncomms3526}{Nat. Commun. {\bf4}, 2526 (2013)}.
   \bibitem{Vie}
   S. Viefers, P. Koskinen, P. S. Deo, and M. Manninen, \href{https://doi.org/10.1016/j.physe.2003.08.076} {Physica E {\bf21}, (2001)}.
   \bibitem{Aha}
   Y. Aharonov and D. Bohm, \href{https://doi.org/10.1103/PhysRev.115.485} {Phys. Rev. {\bf115}, 485 (1959)}.
   \bibitem{Key}
   U. F. Keyser, S. Borck, R.J. Haug, M. Bichler, G. Abstreiter, and W. Wegscheider, \href{https://doi.org/10.1088/0268-1242/17/5/103} {Semicond. Sci. Technol. {\bf17}, L22 (2002)}.
   \bibitem{Pee}
   B. Li and F. M. Peeters, \href{https://doi.org/10.1103/PhysRevB.83.115448}{Phys. Rev. B {\bf83}, 115448 (2011)}.
   \bibitem{Aha1}
   Y. Aharonov and A. Casher, \href{https://doi.org/10.1103/PhysRevLett.53.319} {Phys. Rev. Lett. {\bf53}, 319 (1984)}. 
   \bibitem{Ber}
   T. Bergsten, T. Kobayashi, Y. Sekine, and J. Nitta, \href{https://doi.org/10.1103/PhysRevLett.97.196803}{Phys. Rev. B {\bf97}, 196803 (2006)}.
   \bibitem{Joi}
   F. K. Joibari, Y. M. Blanter, and G. E. W. Bauer, \href{https://doi.org/10.1103/PhysRevB.88.115410}{Phys. Rev. B {\bf88}, 115410 (2013)}.
   \bibitem{Mei}
   Y. Meir, O. E. Wohlman, and Y. Gefen, \href{https://doi.org/10.1103/PhysRevB.42.8351}{Phys. Rev. B {\bf42}, 8351 (1990)}.
   \bibitem{Schi}
   A. Schmid, \href{https://doi.org/10.1103/PhysRevLett.66.80}{Phys. Rev. Lett. {\bf66}, 80 (1991)}
   \bibitem{Tan}
   W. C. Tan and J.C. Inkson, \href{https://doi.org/10.1103/PhysRevB.60.5626}{Phys. Rev. B {\bf60}, 5626 (1999)}.
   \bibitem{Key1}
   U. F. Keyser, C. Fuhner, S. Borck, R. J. Haug, M. Bichler, G. Abstreiter, and W. Wegscheider, \href{https://doi.org/10.1103/PhysRevLett.90.196601}{Phys. Rev. Lett. {\bf90}, 196601 (2003)}.
   \bibitem{Mor}
   A. F. Morpurgo, J. P. Heida, T. M. Klapwijk, B. J. vanWees, \href{https://doi.org/10.1103/PhysRevLett.80.1050}{Phys. Rev. Lett. {\bf80}, 1050 (1999)}.
   \bibitem{Lan}
   M. Buttiker, Y. Imry, R. Landauer, and S. Pinhas, \href{https://doi.org/10.1103/PhysRevB.31.6207}{Phys. Rev. B {\bf31}, 6207 (1985)}.
   \bibitem{Ric}
   D. Frustaglia and K. Richter, \href{https://doi.org/10.1103/PhysRevB.69.235310}{Phys. Rev. B {\bf69}, 235310 (2004)}.
   \bibitem{Ple}
   M Pletyukhov and U. Zulicke, \href{https://doi.org/10.1103/PhysRevB.77.193304}{Phys. Rev. B {\bf77}, 193304 (2008)}.
   \bibitem{Hua1}
   G. Huang, W. Guo, P. Bhattacharya, G. Ariyawansa, and A. G. U. Perera, \href{https://doi.org/10.1063/1.3100407}{Appl. Phys. Lett. {\bf94}, 101115 (2009)}.
   \bibitem{Wu}
   J. Wu, Z. M. Wang, V. G. Dorogan, S. Li, Z. Zhou, H. Li, J. Lee, E. S. Kim, Y. I. Mazur, and G. J. Salamo, \href{ https://doi.org/10.1063/1.4738996}{Appl. Phys. Lett. {\bf101}, 043904 (2012)}.
   \bibitem{You}
   R. J. Young, E. P. Smakman, A. M. Sanchez, P. Hodgson, P. M. Koenraad, and M. Hayne, \href{https://doi.org/10.1063/1.3688037}{Appl. Phys. Lett. {\bf100}, 082104 (2012)}.
   \bibitem{War}
   R. J. Warburton, C. Schaflein, D. Haft, F. Bickel, A. Lorke, K.
Karrai, J. M. Garcia, W. Schoenfeld, and P. M. Petroff, \href{https://doi.org/10.1038/35016030}{Nature (London) {\bf405}, 926 (2000)}.
   \bibitem{Abb}
    M. Abbarchi, C. A. Mastrandrea, A. Vinattieri, S. Sanguinetti, T. Mano, T. Kuroda, N. Koguchi, K. Sakoda, and M. Gurioli, \href{https://doi.org/10.1103/PhysRevB.79.085308}{Phys. Rev. B {\bf79}, 085308 (2009)}.
    \bibitem{FE}
   F. E. Meijer, A. F. Morpurgo, and T. M. Klapwijk, \href{https://doi.org/10.1103/PhysRevB.66.033107}{Phys. Rev. B {\bf66,} 033107 (2002)}.
   \bibitem{Spin_dep1}
J. Nitta, F. E. Meijer, and H. Takayanagi, \href{https://doi.org/10.1063/1.124485}{Appl. Phys. Lett. {\bf 75}, 695 (1999).}

\bibitem{Spin_dep2}
B. Molnar, F. M. Peeters, and P. Vasilopoulos, \href{https://doi.org/10.1103/PhysRevB.69.155335}{Phys. Rev. B {\bf 69}, 155335 (2004).}



\bibitem{Spin_dep4}
P. Foldi, B. Molnar, M. G. Benedict, and F. M. Peeters, \href{https://doi.org/10.1103/PhysRevB.71.033309}{Phys. Rev. B {\bf 71}, 033309 (2005).}


\bibitem{Spin_dep5}
P. Foldi, O. Kalman, M. G. Benedict, and F. M. Peeters,
\href{https://doi.org/10.1103/PhysRevB.73.155325}{Phys. Rev. B {\bf 73}, 155325 (2006)}.

\bibitem{Spin_dep6}
B. Berche, C. Chatelain, and E. Medina,
\href{https://doi.org/10.1088/0143-0807/31/5/026}{Eur. J. Phys. {\bf 31}, 1267 (2010)}.

\bibitem{Graph_exp1}
K. S. Novoselov, A. K. Geim, S. V. Morozov, D. Jiang, Y. Zhang, S. V. Dubonos, I. V. Grigorieva, and A. A. Firsov, 
\href{https://doi.org/10.1126/science.1102896}{Science {\bf 306}, 666 (2004).}

\bibitem{Graph_exp2}
K. S. Novoselov, A. K. Geim, S. V. Morozov, D. Jiang, M. I. Katsnelson, I. V. Grigorieva, S. V. Dubonos, and A. A. Firsov, 
\href{https://doi.org/10.1038/nature04233}{Nature(London) {\bf 438}, 197 (2005).}

\bibitem{Graph_exp3}
Y. Zhang, Y. W. Tan, H. L. Stormer, and P. Kim,
\href{https://doi.org/10.1038/nature04235}{Nature (London) {\bf 413}, 822 (2005).}

\bibitem{Graph_exp4}
V. P. Gusynin and S. G. Sharapov, \href{https://doi.org/10.1103/PhysRevLett.95.146801}{Phys. Rev. Lett. {\bf 95}, 146801 (2005)}.

\bibitem{Zh}
    J. L. Zhu, X. Wang, and N. Yang, \href{https://doi.org/10.1103/PhysRevB.86.125435}{Phys. Rev. B {\bf86}, 125435 (2012)}.

\bibitem{De}
A. De Martino, L. DellAnna, and R. Egger, \href{https://doi.org/10.1103/PhysRevLett.98.066802}{Phys. Rev. Lett. {\bf98}, 066802 (2007)}.
     
\bibitem{Haldane}
F. D. M. Haldane, \href{https://doi.org/10.1103/PhysRevLett.61.2015}{Phys. Rev. Lett. {\bf 61}, 2015 (1988)}.

\bibitem{Grap_Lith1}
S. Russo, J. B. Oostinga, D. Wehenkel, H. B. Heersche, S. S. Sobhani, L. M. K. Vandersypen, and A. F. Morpurgo,
\href{https://doi.org/10.1103/PhysRevB.77.085413}{Phys. Rev. B {\bf 77}, 085413 (2008).}
 
\bibitem{Grap_Lith2}
M. Huefner, F. Molitor, A. Jacobsen, A. Pioda, C. Stampfer, K. Ensslin, and T. Ihn, \href{ https://doi.org/10.1002/pssb.200982284}{Phys. Status Solidi B {\bf 246}, 2756 (2009)}.
   
\bibitem{Grap_Lith3}
M. Huefner, F. Molitor, A. Jacobsen, A. Pioda, C. Stampfer, K. Ensslin, and T. Ihn, 
\href{https://doi.org/10.1088/1367-2630/12/4/043054}{New J. Phys. {\bf 12}, 043054 (2010)}
      \bibitem{PR}
   P. Recher, B. Trauzettel, A. Rycerz, Ya. M. Blanter, C. W. J. Beenakker, and A. F. Morpurgo, \href{https://doi.org/10.1103/PhysRevB.76.235404}{Phys. Rev. B {\bf76,} 235404 (2007)}.
   \bibitem{Fa}
   D. Faria, A. Latge, S. E. Ulloa, and N. Sandler, \href{https://doi.org/10.1103/PhysRevB.87.241403}{Phys. Rev. B {\bf87}, 241403(R) (2013)}.

\bibitem{Bol}
    N. Bolivar, E. Medina, and B. Berche, \href{https://doi.org/10.1103/PhysRevB.89.125413}{Phys. Rev. B {\bf89}, 125413 (2014)}.
    \bibitem{DS}
   D. S. L. Abergel, V. M. Apalkov, and T. Chakraborty, \href{https://doi.org/10.1103/PhysRevB.78.193405}{Phys. Rev. B {\bf78,} 193405 (2008)}.
\bibitem{DR}
   D. R. da Costa, A. Chaves, M. Zarenia, J. M. Pereira Jr., G. A. Farias, and F. M. Peeters, \href{https://doi.org/10.1103/PhysRevB.89.075418}{Phys. Rev. B {\bf89,} 075418 (2014)}.

\bibitem{Graph_Numr1}
P. Recher, B. Trauzettel, A. Rycerz, Ya. M. Blanter, C. W. J. Beenakker, and A. F. Morpurgo, 
\href{https://doi.org/10.1103/PhysRevB.76.235404}{Phys. Rev. B {\bf 76}, 235404 (2007)}


\bibitem{Graph_Numr3}
M. M. Ma, J. W. Ding, and N. Xu, \href{https://doi.org/10.1039/B9NR00044E}{Nanoscale {\bf1}, 387 (2009)}.

\bibitem{Graph_Numr4}
T. Luo, A. P. Iyengar, H. A. Fertig, and L. Brey, \href{https://doi.org/10.1103/PhysRevB.80.165310}{Phys. Rev. B {\bf80}, 165310 (2009)}.

\bibitem{Graph_Numr5}
J. Wurm, M. Wimmer, H. U. Baranger, and K. Richter, \href{https://doi.org/10.1088/0268-1242/25/3/034003}{Semicond. Sci. Technol. {\bf25}, 034003 (2010)}.

\bibitem{Graph_Numr6}
C. H. Yan and L. F. Wei, \href{https://doi.org/10.1088/0953-8984/22/29/295503}{J. Phys.: Cond. Matt. {\bf22}, 295503 (2010)}.
   
\bibitem{Graph_Numr7}
B. L. Huang, M. C. Chang, and C. Y. Mou, \href{https://doi.org/10.1088/0953-8984/24/24/245304}{J. Phys.: Cond. Matt. {\bf24}, 245304 (2012)}.

\bibitem{Graph_Numr9}
A. Lopez, N. Bolivar, E. Medina, and B. Berche, \href{https://doi.org/10.5488/CMP.17.33803}{Cond. Matt. Phys. vol. 17, No. 3, p. 33803: 1-8 (2014)}.

\bibitem{Graph_Numr10}
M. Omidi and E. Faizabadi, \href{https://doi.org/10.1140/epjb/e2014-50607-1}{Eur. Phys. J. B {\bf88}, 30 (2015)}.

\bibitem{Graph_Numr11}
F. R. V. Araujo, D. R. da Costa, A. J. C. Chaves, F. E. B. de Sousa, and J. M. Pereira Jr., \href{https://doi.org/10.1088/1361-648X/ac452e}{J. Phys.: Cond. Matt. {\bf34}, 125503 (2022)}.

\bibitem{Graph_Model1}
M. Zarenia, J. M. Pereira, A. Chaves, F. M. Peeters, and G. A. Farias, \href{https://doi.org/10.1103/PhysRevB.81.045431}{Phys. Rev. B {\bf81}, 045431 (2010)}.
   
\bibitem{Graph_Model2}
M. Zarenia, J. M. Pereira, A. Chaves, F. M. Peeters, and G. A. Farias, \href{https://doi.org/10.1103/PhysRevB.82.119906}{Phys. Rev. B {\bf82}, 119906(E) (2010)}.

\bibitem{Graph_Bilayer}   
M. Zarenia, J. M. Pereira, Jr., F. M. Peeters, and G. A. Farias, \href{https://doi.org/10.1021/nl902302m}{Nano. Lett. {\bf9,} 4088 (2009)}.

\bibitem{hybrid_grapR}
M. Mirzakhani, D. R. da Costa, and F. M. Peeters,
\href{https://doi.org/10.1103/PhysRevB.105.115430}{Phys. Rev. B {\bf 105}, 115430 (2022)}


\bibitem{Graph_Opto}
M. Samal, N. Barange, D. -H. Ko, and K. Yun, \href{https://doi.org/10.1021/acs.jpcc.5b05225}{J. Phys. Chem. C {\bf 119}, 19619 (2015)}.
       
\bibitem{Graph_Interf}
D. J. P. de Sousa, A. Chaves, J. M. Pereira Jr., and G. A. Farias, \href{https://doi.org/10.1063/1.4973902}{J. Appl. Phys. {\bf 121}, 024302 (2017)}.
   
   \bibitem{MI}
   \href{https://doi.org/10.1017/CBO9781139031080}{M. I. Katsnelson, Graphene: Carbon in Two Dimensions (Cambridge University Press, Cambridge, 2012)}.
   
   \bibitem{Su}
   B. Sutherland, \href{https://doi.org/10.1103/PhysRevB.34.5208}{Phys. Rev. B {\bf34}, 5208 (1986)}.

   \bibitem{Vi}
   J. Vidal, R. Mosseri, and B. Doucot, \href{https://doi.org/10.1103/PhysRevLett.81.5888}{Phys. Rev. Lett. {\bf81}, 5888(1998)}.
   
   \bibitem{Wa}
   F. Wang and Y. Ran, \href{https://doi.org/10.1103/PhysRevB.84.241103}{Phys. Rev. B {\bf84}, 241103(R) (2011)}.
   
   \bibitem{Ur}
    D. F. Urban, D. Bercioux, M. Wimmer, and W. Husler, \href{https://doi.org/10.1103/PhysRevB.84.115136}{Phys. Rev. B {\bf84}, 115136 (2011)}.
    
    \bibitem{Mal}
     J. D. Malcolm and E. J. Nicol, \href{https://doi.org/10.1103/PhysRevB.92.035118}{Phys. Rev. B {\bf92}, 035118 (2015)}.

\bibitem{T3_Hall1}
E. Illes, J. P. Carbotte, and E. J. Nicol, \href{http://dx.doi.org/10.1103/PhysRevB.92.245410}{Phys. Rev. B {\bf 92}, 245410 (2015)}.

\bibitem{T3_Hall2}
T. Biswas and T. K. Ghosh, \href{https://doi.org/10.1088/0953-8984/28/49/495302}{J. Phys.: Condens. Matter {\bf 28}, 495302 (2016)}.

\bibitem{Klein2}
E. Illes and E. J. Nicol, \href{https://doi.org/10.1103/PhysRevB.95.235432}{Phys. Rev. B {\bf 95}, 235432 (2017)}.

\bibitem{Weiss}
SK. F. Islam and P. Dutta, \href{https://doi.org/10.1103/PhysRevB.96.045418}{Physical Review B {\bf 96}, 045418 (2017)}.

\bibitem{ZB}
T. Biswas and T. K. Ghosh, 
\href{https://doi.org/10.1088/1361-648X/aaa60b}{J. Phys.: Condens. Matter {\bf 30}, 075301 (2018)}.

\bibitem{Plasmon1}
J. D. Malcolm and E. J. Nicol, \href{https://doi.org/10.1103/PhysRevB.93.165433}{Phys. Rev. B {\bf 93}, 165433 (2016)}.

\bibitem{Plasmon2}
A. Balassis, D. Dahal, G. Gumbs, A. Iurov, D. Huang and O.
Roslyak, 
\href{https://doi.org/10.1088/1361-648X/aba97f}{J. Phys.: Condens. Matter {\bf 32}, 485301 (2020)}.

\bibitem{Plasmon3}
A. Iurov, G. Gumbs, and D. Huang,
\href{https://doi.org/10.1088/1361-648X/ab9bcb}{J. Phys.: Condens. Matter {\bf 32}, 415303 (2020)}. 

\bibitem{Plasmon4}
A. Iurov, L. Zhemchuzhna, G. Gumbs, D. Huang, D. Dahal, and Y. Abranyos,
\href{https://doi.org/10.1103/PhysRevB.105.245414}{Phys. Rev. B {\bf 105}, 245414 (2022)}.


\bibitem{Mag_Opt1}
E. Illes and E. J. Nicol,
\href{https://doi.org/10.1103/PhysRevB.94.125435}{Phys. Rev. B {\bf 94}, 125435 (2016)}.

\bibitem{Mag_Opt2}
A. D. Kovacs, G. David, B. Dora, and J. Cserti, \href{https://doi.org/10.1103/PhysRevB.95.035414}{Phys. Rev. B {\bf 95}, 035414 (2017)}.

\bibitem{Mag_Opt3}
Y. R. Chen, Y. Xu, J. Wang, J. F. Liu, and Z. Ma, 
\href{https://doi.org/10.1103/PhysRevB.99.045420}{Phys. Rev. B {\bf 99}, 045420 (2019)}.

\bibitem{Mag_Opt4}
L. Chen, J. Zuber, Z. Ma, and C. Zhang,
\href{https://doi.org/10.1103/PhysRevB.100.035440}{Phys. Rev. B {\bf 100}, 035440 (2019)}.



\bibitem{RKKY1}
D. O. Oriekhov and V. P. Gusynin, \href{https://doi.org/10.1103/PhysRevB.101.235162}{Phys. Rev. B {\bf 101}, 235162 (2020)}.

\bibitem{RKKY2}
O. Roslyak, G. Gumbs, A. Balassis, H. Elsayed, \href{https://doi.org/10.1103/PhysRevB.103.075418}{Phys. Rev. B {\bf 103}, 075418 (2021)}.

\bibitem{Min_Con}
J. Wang, J. F. Liu, and C. S. Ting,
\href{https://doi.org/10.1103/PhysRevB.101.205420}
{Phys. Rev. B {\bf 101}, 205420 (2020)}.

\bibitem{Ghosh_Topo}
B. Dey, P. Kapri, O. Pal, and T. K. Ghosh,
\href{https://doi.org/10.1103/PhysRevB.101.235406}{Phys. Rev. B {\bf 101}, 235406 (2020)}.

\bibitem{spin_hall}
J. Wang and J. F. Liu,
\href{https://doi.org/10.1103/PhysRevB.103.075419}{Phys. Rev. B {\bf 103}, 075419 (2021)}.


\bibitem{aT3_Mijanur}
M. Islam and P. Kapri,
\href{https://doi.org/10.1088/1361-648X/acae13}{J. Phys.: Condens. Matter {\bf 35}, 105301 (2023)}.
 
\bibitem{Bashab1}
B. Dey and T. K. Ghosh,
\href{https://doi.org/10.1103/PhysRevB.98.075422}{Phys. Rev. B {\bf 98}, 075422 (2018)}.

\bibitem{Bashab2}
B. Dey and T. K. Ghosh, \href{https://doi.org/10.1103/PhysRevB.99.205429}{Phys. Rev. B {\bf 99}, 205429 (2019)}.

\bibitem{Iurov_Floq}
A. Iurov, G. Gumbs, and D. Huang,
\href{https://doi.org/10.1103/PhysRevB.99.205135}{Phys. Rev. B {\bf 99}, 205135 (2019)}.


\bibitem{Mojarro}
M. A. Mojarro, V. G. Ibarra-Sierra, J. C. Sandoval-Santana, R. Carrillo-Bastos, and G. G. Naumis, \href{https://doi.org/10.1103/PhysRevB.101.165305}{Phys. Rev. B {\bf 101}, 165305 (2020).}



\bibitem{TB_Floq}
L. Tamang, T. Nag, and T. Biswas,
\href{https://doi.org/10.1103/PhysRevB.104.174308}{Phys. Rev. B {\bf 104}, 174308 (2021)}.


\bibitem{Trans_aT3}
Z. P. Niu and S. Jun Wang,
\href{https://doi.org/10.1088/1361-6463/ac5992}{J. Phys. D: Appl. Phys. {\bf 55}, 255303 (2022)}. 
  
\bibitem{TB_Trans}
L. Tamang and T. Biswas,
\href{https://doi.org/10.1103/PhysRevB.107.085408}{Phys. Rev. B {\bf 107}, 085408 (2023)}.     
    
\bibitem{Sayan_dice}
S. Mondal and S. Basu, \href{https://doi.org/10.1103/PhysRevB.107.035421}{Phys. Rev. B {\bf 107}, 035421 (2023)}. 

\bibitem{Rahul1}
R. Soni, N. Kaushal, S. Okamoto, and E. Dagotto, \href{https://doi.org/10.1103/PhysRevB.102.045105}{Phys. Rev. B {\bf 102}, 045105 (2020)}.

\bibitem{Rahul2}
R. Soni, A. B. Sanyal, N. Kaushal, S. Okamoto, A. Moreo, and E. Dagotto, \href{https://doi.org/10.1103/PhysRevB.104.235115}{Phys. Rev.B {\bf 104}, 235115 (2021)}.
    
\bibitem{QSH_alpha}
J. Wang and J. F. Liu, \href{https://doi.org/10.1103/PhysRevB.103.075419}{Phys. Rev. B {\bf103}, 075419 (2021)}.

\bibitem{Kane}
C. L. Kane and E. J. Mele, \href{https://doi.org/10.1103/PhysRevLett.95.226801}{Phys. Rev. Lett. {\bf 95}, 226801 (2005)}.

\bibitem{Func}
G. Autes and O. V. Yazyev, \href{https://doi.org/10.1103/PhysRevB.87.241404}{Phys. Rev. B {\bf 87}, 241404 (2013)}.

\bibitem{Just1}
H. Min, J. E. Hill, N. A. Sinitsyn, B. R. Sahu, L. Kleinman, and A. H. MacDonald, \href{https://doi.org/10.1103/PhysRevB.74.165310}{Phys. Rev. B {\bf 74}, 165310 (2006)}.

\bibitem{Just2}
Kh. Shakouri, B. Szafran, M. Esmaeilzadeh, and F. M. Peeters, \href{https://doi.org/10.1103/PhysRevB.85.165314}{Phys. Rev. B {\bf 85}, 165314 (2012)}.

\bibitem{Just3}
L. Brey and H. A. Fertig, \href{https://doi.org/10.1103/PhysRevB.73.235411}{Phys. Rev. B {\bf 73}, 235411 (2006)}.

\bibitem{Mijanur}
M. Islam, T. Biswas, and S. Basu, \href{https://doi.org/10.48550/arXiv.2304.08830}{	arXiv:2304.08830 (2023)}.

\bibitem{spin_TMD}
X. Xu, W. Yao, D. Xiao, and T. F. Heinz, \href{https://doi.org/10.1038/nphys2942}{Nat. Phys. {\bf10}, 343 (2014)}.

\bibitem{spin-Silicene}
M. Ezawa \href{https://doi.org/10.1103/PhysRevB.86.161407}{Phys. Rev. B {\bf86}, 161407(R) (2012)}.

\bibitem{Bles}
A. C. B. Jayich, W. E. Shanks, B. Peaudecerf, E. Ginossar, F. V. Oppen, L. Glazman, and J. G. E. Harris, \href{https://doi.org/10.1126/science.1178139}{Science {\bf326},272 (2009)}.

\bibitem{Ci2}
L. Ci, Z. Xu, L. Wang, W. Gao, F. Ding, K. F. Kelly, B. I. Yakobson, and P. M. Ajayan, \href{https://doi.org/10.1007/s12274-008-8020-9}{Nano Res. {\bf1}, 116 (2008)}. 

\bibitem{Bala}
J. Balakrishnan, G. K. W. Koon, M. Jaiswal, A. H. Castro Neto, and B. Ozyilmaz,  \href{https://doi.org/10.1038/nphys2576}{Nat. Phys. {\bf9}, 284 (2013)}.

\bibitem{Butt}
M. Buttiker, Y. Imry, and R. Landauer, \href{https://doi.org/10.1016/0375-9601(83)90011-7}{Phys. Lett. A {\bf96}, 365 (1983)}.

\end{thebibliography}
\end{document}